\documentclass[aps,pra,reprint,superscriptaddress,twocolumn]{revtex4-2}
\usepackage{amssymb}		
\usepackage{amsmath}
\usepackage{amsthm}
	\newtheorem{myTheorem}{Theorem}
\usepackage{graphicx}
	\graphicspath{{Figures/}}
\usepackage[linktoc=all]{hyperref}
	\hypersetup{final=true,hidelinks}
\usepackage{float}
\usepackage{adjustbox}

\newcommand{\ket}[1]{{| #1 \rangle}}

\begin{document}

\title{Boson Sampling in Low-depth Optical Systems}
\author{R.~van~der~Meer}
\email[]{r.vandermeer-1@utwente.nl}
\affiliation{Adaptive Quantum Optics, MESA+ Institute for Nanotechnology, University of Twente, PO Box 217, 7500 AE Enschede, The Netherlands}

\author{S. Huber}
\affiliation{Leibniz Supercomputing Centre of the Bavarian Academy of Sciences and Humanities, Boltzmannstr. 1, 85748 Garching, Germany}

\author{P. W. H. Pinkse}
\affiliation{Adaptive Quantum Optics, MESA+ Institute for Nanotechnology, University of Twente, PO Box 217, 7500 AE Enschede, The Netherlands}

\author{R. García-Patrón}
\affiliation{School of Informatics, University of Edinburgh, Edinburgh EH8 9AB, UK}
\author{J. J. Renema}
\affiliation{Adaptive Quantum Optics, MESA+ Institute for Nanotechnology, University of Twente, PO Box 217, 7500 AE Enschede, The Netherlands}

\date{\today}

\begin{abstract}
\textbf{\noindent Optical losses are the main obstacle to demonstrating a quantum advantage via boson sampling without leaving open the possibility of classical spoofing. We propose a method for generating low-depth optical circuits suitable for boson sampling with very high efficiencies. Our circuits require only a constant number of optical components (namely three) to implement an optical transformation suitable for demonstrating a quantum advantage. Consequently, our proposal has a constant optical loss regardless of the number of optical modes. We argue that sampling from our family of circuits is computationally hard by providing numerical evidence that our family of circuits converges to that of the original boson sampling proposal in the limit of large optical systems. Our work opens a new route to demonstrate an optical quantum advantage.}
\end{abstract}

\maketitle
Recently, the world has seen the first claims of a demonstration of a quantum advantage \cite{arute_2019_Nature, zhong_2020_Science,zhong_2021_ArXiv,wu_2021_ArXiv}, i.e. the first systems in which a quantum computer outperforms a classical one at some specific, well-defined computational task. These systems represent a major milestone on the road toward applications of quantum computing for tasks beyond those that can be carried out with classical computers. 

Boson sampling is a protocol to demonstrate a quantum advantage using quantum interference in linear optics \cite{aaronson_2013_TheoryComput}. In boson sampling, $n$ indistinguishable photons are created at photon sources and sent through a linear optical network whose transformation matrix is chosen uniformly according to the Haar measure. The task is to generate samples from the output distribution of this interferometer when measured in the Fock basis. For a sufficiently large optical system, the probability distribution is given by permanents of matrices that are very well approximated by identical and independently distributed (i.i.d.) Gaussian elements. This distribution is believed to be hard to sample from for a classical computer, with the computational cost scaling as $n 2^n$ for the best-known classical algorithms \cite{ryser_1963_}, which places the boundary of a quantum advantage at $n \approx 50$ photons \cite{wu_2018_Natl.Sci.Rev.}. At the same time, a linear optical network can natively implement this computational task through multi-photon interference and thereby generate samples from the target output distribution efficiently. 

Since this protocol was proposed in 2011, there has been major progress on its experimental implementation. This started with smaller-scale boson sampling experiments \cite{spring_2013_Science,broome_2013_Science,tillmann_2013_NatPhotonics,crespi_2013_NatPhotonics,loredo_2017_PRL,wang_2019_Phys.Rev.Lett.,he_2017_PhysRevLett,zhong_2018_PRL} to the current state of the art experiments \cite{lund_2014_Phys.Rev.Lett.,hamilton_2017_Phys.Rev.Lett.}, which consists of a $144$ mode Gaussian boson sampler with 25 two-mode squeezed photon sources \cite{zhong_2021_ArXiv}. 

The challenge in carrying out these experiments is to adhere as closely as possible to the original boson sampling protocol, and not introduce any imperfections, which may give rise to classical simulability. The general physical picture is that noise degrades the level of quantum interference in these systems, opening up the way for either a direct simulation \cite{renema_2018_Phys.Rev.Lett.,renema_2018_ArXiv,garcia-patron_2019_Quantum,brod_2020_Quantum,renema_2020_Phys.Rev.A,2109.11525,shi_2021_ArXiv}, or spoofing a benchmark test \cite{pan_2021_ArXiv,bulmer_2021_ArXiv}.

The strongest imperfection in large-scale boson sampling experiments is photon loss, which is an imperfection where not every photon that is generated in the sources ultimately results in a detection event in one of the detectors. For all proposed variants of boson sampling, loss is an imperfection that degrades the strength of quantum interference, although only for some variants it is known how to use this fact to construct an efficient classical simulation \cite{renema_2020_Phys.Rev.A}. Finding ways of improving the optical transmission of boson sampling experiments is therefore an urgent problem. This is especially true since, in all current implementations of boson sampling, the number of optical components that must be traversed grows linearly with the width of the optical circuit, and hence the optical loss grows exponentially with the circuit size \cite{GarcaPatrn2019}.

Another imperfection that plays a role in large-scale boson sampling experiments is detector blinding, which occurs when the number of modes $m$ is comparable to the number of photons $n$, meaning that a significant fraction of photons will end up in identical output modes, which is termed a \textit{collision}. In this situation, if threshold detectors are used at the outputs, some photons are not detected, which constitutes an imperfection that can be used for simulation \cite{popova_2021_ArXiv}. In order to avoid collisions, the number of optical modes must scale as $m \propto n^2$, a scaling that is typically not obeyed in experiments \cite{shchesnovich_2020_Int.J.QuantumInform.,zhong_2020_Science,zhong_2021_ArXiv}.

\begin{figure}
    \includegraphics{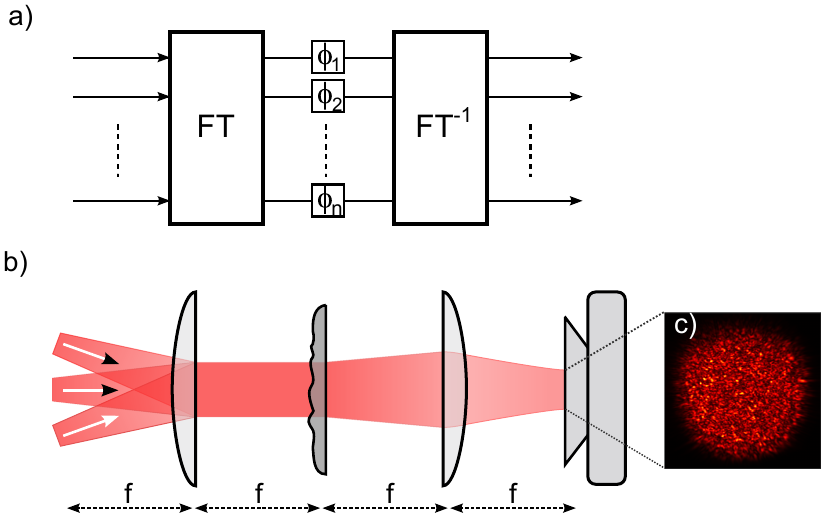}
    \caption{A schematic of the proposed network with 4 input modes. a) The mathematical description of the physical apparatus as shown in b). The apparatus consists of two lenses and a thin, rough surface in between the lenses. The random interference pattern as observed on the detector (c) is known as speckle.}
    \label{fig:figSchematic}
\end{figure}

In this work, we present a simultaneous solution to the problems of optical loss and detector blinding by proposing a constant-optical depth circuit that is scalable to the required thousands of optical modes. This circuit is realized in the form of a thin optical scatterer (e.g. a ground glass plate) placed in a $4f$-configuration in the simultaneous focal point of two lenses (see Fig. \ref{fig:figSchematic}). We make use of the fact that a lens implements a Fourier transform on an almost arbitrary number of optical modes - up to $1$ million modes per $\rm{mm}^2$ \cite{mosk_2012_NatPhoton} - to conveniently realize a network with three components and constant optical loss in the system size. When a Fourier transform is implemented as pairwise mode interactions, e.g. in integrated optics, the number of required interactions per mode in the circuit scales as the logarithm of the number of modes involved. \cite{barak_2007_J.Opt.Soc.Am.B,crespi_2016_NatCommun}.
We estimate that the overall efficiency of this protocol will be $85-90\%$, which would be sufficient to defeat all currently known classical simulation algorithms for boson sampling.

The price of this low-tech implementation is a complete loss of reconfigurability of the optical system. Furthermore, the fact that our scattering system scrambles the phase of each incoming k-vector in an identical way induces an additional constraint to the transmission matrix $U$, namely that it must be of \textit{circulant} form, i.e. that each row of the sampling matrix is a copy of the previous row shifted by one entry. These additional correlations in the sampling matrix could \textit{prima facie} affect the computational hardness of the protocol. 
The bulk of this work is concerned with presenting a combination of numerically and theoretical results which provide evidence that these additional correlations disappear in the appropriate limits of large numbers of photons and modes, thereby making our system suitable as a low-loss implementation of boson sampling. We name our proposal \textit{circulant boson sampling} after the circulant nature of the scattering matrix. 

Our work is organized as follows: we start in Section \ref{secScatter} with some physical arguments based on concepts from scattering theory. Then, Section. \ref{secPrelim} formally introduces the circulant boson sampling and identifies the problems which need to be solved to guarantee the computational complexity. Section \ref{secOverview} gives an informal overview of the main results on demonstrating the computational complexity. Sections \ref{secIID} and \ref{secNoCol} show the detailed results in which we show how circulant boson sampling maintains computational complexity. Finally, Section \ref{secHardware} addresses the technical feasibility of the system and focuses on the required hardware.

\section{Thin scattering systems}\label{secScatter}
 In this section, we will use some concepts from scattering theory to build up physical intuition for circulant boson sampling. For this, we start with a short description of thick scattering systems and then introduce the circulant network. 
 
 Perhaps counterintuitively, disorder is a desirable feature in many classical optical systems, where a high number of independent optical modes is needed. An example is in microscopy, where a diffusor is used to convert a point source into quasi-uniform illumination. Functionalizing disorder to achieve new optical applications is an active field of research \cite{vellekoop_2007_OptLett,mosk_2012_NatPhoton,rotter_2017_Rev.Mod.Phys.}. 
 
 A crucial factor that determines the properties of a scattering system is how often light is scattered while traversing the medium. Everyday optical systems in which scattering plays a role are typically multiple-scattering (`thick') systems, examples of which are white paint and milk. In multiple-scattering materials, incident light scatters many times on average before it diffuses out of the material. Repeated scattering events collectively scramble the polarization of the incoming light and effectively erase all directionality, i.e., the light is scattered in all $4\pi$ directions. 
 However, even in thick optical samples, the output distribution pattern is not uniform when pumped with coherent light. The random paths which the light takes through the sample are coherent and hence undergo interference. This random interference leads to a pattern of spatial intensity variations, which is referred to as speckle. An example of speckle is shown in Fig. \ref{fig:figSchematic}b).

The properties of speckle patterns have been extensively studied \cite{beenakker_2009_RMT,goodman2007speckle,mosk_2012_NatPhoton}. A key property of speckle is its high degree of randomness. Random interference of many independent optical paths causes a complex Gaussian distribution of the elements of the scattering matrix. This is observed experimentally through the Rayleigh distribution of the optical field strengths leading to an exponential distribution of the intensities of the speckle pattern \cite{goodman2007speckle,beenakker_2009_RMT,dilorenzopires_2012_Phys.Rev.A}. 
In the limit of thick optical systems, small submatrices of the scattering matrix approach an i.i.d. Gaussian distribution, as the scattering effectively erases all structure and only a part of all transmission modes can be observed. The fact that the matrix elements are i.i.d. Gaussian distributed turns out to be precisely the requirement for the computational complexity of boson sampling \cite{aaronson_2013_TheoryComput}.

Unfortunately, boson sampling also requires a low-loss optical system, and for this, thick scattering systems are not suitable \cite{defienne_2014_Opt.Lett.a,huisman_2014_ApplPhysB,wolterink_2016_PhysRevA}. The ultimate reason for this is that in a multiple-scattering system the light is scattered over a $4\pi$ solid angle, which makes a collection of all the light scattered from the sample experimentally infeasible.

The network proposed in Fig. \ref{fig:figSchematic}, which uses a single thin scattering layer,  can be thought of as a system that combines the high degree of randomness of a thick scattering system with low optical loss. Such a system still grants access to a large number of modes, but at the same time reduces the losses since all the light will be transmitted over a small spread of spatial angles \cite{goodman2007speckle}. 

Unfortunately, having only one layer of scatterers does not allow for the same degree of mode mixing as is the case in thick scattering media. The consequence of this is that while the individual matrix elements in the scattering matrix all follow a Gaussian distribution (in the language of scattering theory `exhibit fully developed speckle') \cite{peeters_2010_Phys.Rev.Lett.}, 
they exhibit significant correlations (in the language of scattering theory: they strongly exhibit a memory effect), and it is not  given that the distribution of submatrices of such a matrix approaches an i.i.d. Gaussian distribution. It is precisely these additional circulant correlations that are the subject of this work.

\section{Boson Sampling \label{secPrelim}}
In this section, we will first introduce boson sampling. Then, we will introduce a few key arguments from its original proposal \cite{aaronson_2013_TheoryComput} regarding the hardness of boson sampling, followed by our proposal for a modified protocol based on circulant matrices. A comprehensive and more in-depth discussion on boson sampling can be found in \cite{brod_2019_AP}.

In a boson sampling experiment, $n$ indistinguishable photons are injected in $n$ distinct input modes of an $m\times m$ random lossless interferometer ($m > n$). The photons interfere as they propagate through the interferometer, resulting in an output probability distribution comprised of $\binom{m+n-1}{n}$ entries. 

Assume without loss of generality that the photons are incident on the first $n$ input modes of the system, such that the input wave function is given by:
\begin{equation}
    \ket{\psi_{\rm in}} = \ket{1}^{\bigotimes n}\ket{0}^{\bigotimes m-n},
    \label{eq:eqInput}
\end{equation}
and let the transmission through the interferometer be described by a complex unitary matrix $U \in \mathbb{C}^{m\times m}$ drawn from the Haar measure. The output state is then given by:
\begin{equation}
    \ket{\Psi_{\textrm{out}}} = \sum_{\rm {s}} P_{\rm{s}} \ket{n_{o,1}^{(\rm{s})}, n_{o,2}^{(\rm{s})},...,n_{o,m}^{(\rm{s})}},
\end{equation}
with $s$ a particular output configuration and $P_{\rm{s}}$ its corresponding probability amplitude \cite{scheel_2004_ArXiv}. 
Finally, $n_{o,m}^{(\rm{s})}$ denotes the number of bosons in the output channel $m$ given this configuration $\rm{s}$. The probability associated with a certain outcome $P_{\rm{s}}$ is given by the interference of how the $n$ input photons can end up at that specific output configuration $\rm{s}$. This is given by the permanent of an $n\times n$ submatrix $U_{\rm{T}}$ relating the used input modes to the output configuration \cite{aaronson_2013_TheoryComput, scheel_2004_ArXiv}
\begin{equation}
    P_{\rm{s}} = \frac{ {\rm perm}(U_{\rm T})}{\sqrt{n_{o1}^{(\rm{s})}!,...,n_{om}^{(\rm{s})}!}},
\end{equation}
with the submatrix $U_{\rm T} = U_{\rho,s}$, with $\rho$ the input configuration as defined in Eq. \ref{eq:eqInput}. The denominator is a normalization term and perm denotes the permanent. The permanent is defined as:
\begin{equation}
    \rm{perm}(U) = \sum_{\sigma \in S_n} \prod_{j=1}^n{U_{j,\sigma(j)}},
\end{equation}
with $\sigma$ summing over the set of all permutations $S_n$.

The computational complexity of boson sampling ultimately derives from the conjecture that approximate sampling over the probability distribution given by $|P_{\rm{s}}|^2$ is computationally hard, if $U_{\rm{T}}$ is given by a matrix of i.i.d. Gaussian entries. The intuition behind the choice of this family of matrices is that i.i.d. Gaussian matrices have no internal structure which a classical algorithm could exploit to achieve a speedup.

For the matrices $U_{\rm{T}}$ to be close to i.i.d. Gaussian, two conditions must hold. First, any $n$-by-$n$ submatrix of $U$ must converge to a matrix of i.i.d. Gaussians in the limit of a large number of modes. The intuition here is that as the submatrix becomes smaller with respect to the overall matrix $U$, the unitary constraint becomes less dominant, removing any correlations between the matrix elements. Aaronson and Arkhipov were able to show that if $U$ is drawn uniformly according to the Haar measure, then for a scaling of $m \propto n^5 \rm{log}\,n$, an $n$-by-$n$ submatrix is close to i.i.d. Gaussian.

The second requirement is that the output distribution must actually be made up of $n$-by-$n$ matrices drawn from $U$, avoiding repeated rows or columns in $U_{\rm{T}}$. These arise when photons emerge from the same optical mode, which is termed a \textit{collision}. The probability of collisions must be made arbitrarily small, a condition which is satisfied for Haar-random matrices when $m \propto n^2$. 

To argue the hardness of our proposed protocol, we must show that these two conditions hold (perhaps with different polynomial scaling in the number of modes) for our proposed family of sampling matrices. Our goal in what follows will be to provide evidence that the intuition from sampling over Haar-random matrices holds also for circulant matrices, namely that the additional correlations which are introduced by the constraints on the matrices decay when the number of modes is made large enough with respect to the number of photons.

In our protocol, the only change we make compared to the original proposal \cite{aaronson_2013_TheoryComput} is the family of optical circuits over which we sample. Instead of Haar-random matrices, we now sample over matrices of the form $U_{\rm c} = F \Phi F^{-1}$, where $F$ is an $n$-by-$n$ Fourier matrix with the appropriate normalization and $\Phi$ is a diagonal matrix whose entries are given by $\Phi_{jj} = \exp{(i\phi_j)}$, with each $\phi_{j}$ chosen independently on the interval $[0,2\pi]$. The resulting matrix has the form: 
\begin{equation}
    \begin{aligned}
        U_{c} = \begin{bmatrix} &c_1 &c_2 &\ldots &c_m\\ &c_m &c_1 &\ldots &c_{m-1}\\ &\vdots &\vdots &\ddots &\vdots\\ &c_2 &\ldots &\ldots &c_1 \end{bmatrix},
    \end{aligned}
\end{equation}
with $c_{j} \in \mathbb{C}$, and is called a circulant matrix. Note that this matrix has only $m$ unique elements since each row or column is a repetition of the previous one where all elements have been shifted by one position. These additional correlations are an additional obstacle in letting the elements of a truncated matrix $U_{\rm T}$ approach i.i.d. Gaussians.

\section{Overview of results \label{secOverview}}
In this section, we provide a general overview of our results. Our strategy to argue the computational complexity of circulant boson sampling is to closely follow the proof for standard boson sampling \cite{aaronson_2013_TheoryComput}. To this end, we must show that (1) truncations of circulant matrices should approach i.i.d. Gaussians with polynomial scaling in $n$ and $m$ and (2) that there is a polynomial scaling between $n$ and $m$ in which the no-collision condition holds. In this section we provide an informal overview of the main results on these two points. For both points, we rely on a combination of numerical and analytic results to provide our evidence. 

A crucial fact about our system is that the output configurations are not equiprobable, unlike the Haar-random case \cite{arkhipov_2012}. This is due to the fact that repeated entries in the sampling matrix give rise to higher order moments of matrix elements in the expected value of the permanent. This fact means that the symmetry arguments employed in \cite{aaronson_2013_TheoryComput, arkhipov_2012} are not available to us.

\textbf{Truncations - analytical result}
We show that the variation distance $\Delta_{\rm C}$ between the probability distribution of submatrices of circulants and the i.i.d. Gaussian ensemble is upper bounded by
\begin{equation}
    \Delta_{\rm C} \leq \acute{C} \frac{n^2}{\sqrt{m}}
\end{equation}
For some universal constant $\acute{C}>0$.

\textbf{Truncations - numerical results} We provide evidence that the eigenvalue spectrum of a truncated circulant matrix strongly resembles that of a truncated Haar-random matrix. In particular, we provide numerical evidence that the fidelity between the eigenvalue distribution of a Gaussian and truncated circulant matrix remains constant at $m \propto n$ scaling. 

Furthermore, we provide evidence that the convergence of permanents of truncated circulant matrices to permanents of i.i.d. Gaussian matrices have the same scaling as the convergence of permanents of truncated Haar matrices to permanents of i.i.d. Gaussian matrices. This is another, independent piece of evidence for the convergence of submatrices of circulant matrices to i.i.d. Gaussians.

\textbf{No collisions - Analytical results}
Collisions in the matrix elements reduce the computational complexity of boson sampling. The original proposal uses Haar-random matrices, where every photon `blocks off' a single potential output mode for the remaining photons (namely the one which it occupies). In circulant boson sampling, the cyclic nature of circulant matrices means that each photon blocks off $n$ adjacent modes as well.

We show that if we assume all outcomes in the boson sampler are equiprobable averaged over all circulant matrices (which is only approximately true for circulant matrices), a scaling of $m \propto n^3$ is required to achieve a constant probability of an outcome with repeated elements in the matrix (i.e. a collision).

\textbf{No collisions - numerical results} We use a Monte Carlo simulation to provide evidence that the $m \propto n^3$ scaling inferred from our simplified calculation assuming equiprobable outcomes also holds when we take into account the fact that not all outcomes are equiprobable.

\section{Truncations approach i.i.d. Gaussian \label{secIID}}
In this section, we provide evidence that sufficiently small truncations of circulant matrices become close, in variation distance, to a matrix with i.i.d. Gaussian elements. To this end, we follow the strategy of the original boson sampling proposal \cite{aaronson_2013_TheoryComput}. Their approach has two steps. Firstly, they showed that a polynomially small submatrix of a larger, Haar-random unitary matrix becomes approximately Gaussian distributed. Then this result was used to prove the stronger statement, namely the multiplicative bound
\begin{equation}
    p_S(X) \leq (1 + O(\delta))p_G(X),
\end{equation}
$\forall X \in C^{n\times n}$ and with $p_S$ and $p_G$ the probability density functions of Haar random unitaries and i.i.d. Gaussians, respectively.

In the following subsections, we will first show that the submatrix of a circulant matrix approaches the i.i.d. Gaussian ensemble additively. Unfortunately, we were unable to prove the multiplicative bound from the additive bound. As additional evidence of convergence, we study both the eigenvalue spectra of truncated circulant matrices and the average outcome of permanents of truncated circulant matrices. Both approaches independently provide insight to the convergence to i.i.d. Gaussians.

\subsection{Analytical results \label{ssecWeak}}
We first present our analytical results providing evidence that the distribution of $n$-by-$n$ submatrices of non-identical elements (i.e. without repetitions in the elements) converges to an i.i.d. Gaussian distribution. 

Rather than looking at a submatrix, we can limit ourselves to a single vector containing all elements of the submatrix. The vector has a length of $d=m^2$ and thus corresponds to the submatrix. The goal is to show that the effect of the unitary constraint on this vector becomes negligible for some suitable scaling of the length of the vector. For this first step, we use theorem $1.1$ of \cite{bentkus_2005_TheoryProbab.Appl.}.

\begin{myTheorem}
Let $\Re^d$ be a real Euclidian d-dimensional space of vectors $x=(x_1,\ldots,x_d)$ with the norm $||x||^2_2=\sum_{i=1}^d x_i^2$ and the scalar product $\left\langle x,x \right\rangle = ||x||^2_2$. Let $X_1,\ldots,X_n$ be independent random vectors with a common mean $E(X) = 0$. Write $S=\sum_{j=0}^n X_j$. Throughout we assume that S has a degenerated distribution in the sense that the covariance operator, say $\Sigma = Cov(S)$, is invertible. Let $Z \sim N(0,\Sigma)$ be a Gaussian random vector with the same covariance matrix $S$. Let $C$ be the class of all convex subsets of $\Re^d$. Then there exists a universal constant $C>0$ such that
\begin{equation}
    \begin{aligned}
    \Delta_{\rm C} := &\sup_{A\in C} | Pr\{ S\in A\} -Pr\{Z \in A\}| \\
    \leq &C d^{\frac{1}{4}} \sum_{j=1}^n E(||\Sigma^{-\frac{1}{2}}X_j||^3_2),
    \end{aligned}
\end{equation}
where $\Sigma^{-\frac{1}{2}}$ stands for the positive root of $\Sigma$.
\end{myTheorem}

We now show how we can apply this Theorem to the case where we pick $d$ elements of a vector of length $n$ which is the Fourier transform of a vector of random phases. Let $\Phi_1,\ldots,\Phi_m$ be i.i.d. uniform on $\left[0,2\pi\right]$. Fix $k_1,\ldots,k_n \in \{1,\ldots,m\}$. For $j=0,\ldots,m-1$, define $X_j^{(n)}$ to be the vector containing all matrix elements
\begin{equation}
\begin{aligned}
    X_j^{(n)} &= \frac{1}{\sqrt{n}}\begin{bmatrix} w_n^{jk_1}e^{i\phi_j}\\\vdots\\w_n^{jk_d}e^{i\phi_j} \end{bmatrix}\\
    &= \frac{1}{n}\begin{bmatrix} \cos(\phi_j +2\pi\frac{jk_1}{n})\\ \sin(\phi_j +2\pi\frac{jk_1}{n}) \\
    \cos(\phi_j +2\pi\frac{jk_2}{n})\\ \vdots\\ \sin(\phi_j +2\pi\frac{jk_n}{n}) \end{bmatrix} \in \Re^{2d},
\end{aligned}
\end{equation}
where the first line is a column vector whose elements are given by a discrete Fourier transform applied on the vector of phases $\phi_j$, and $w_n^{jk_l} = \frac{1}{\sqrt{n}} \exp{(\frac{-2\pi i}{m}k_l l}$). The second line expands the space by a factor two by splitting up the real and imaginary part of the previous line.

The $X_j^{(n)}$ are independent and clearly, $E(X_j^{(n)})=0$ for all $j=0,\ldots,n-1$. Furthermore, it can be checked that (under some assumptions on $k_1,\ldots,k_d$, see \cite{meckes_2009_})
\begin{equation}
    Cov( \sum_{j=0}^{n-1} X_j^{(n)}) = \frac{1}{2} I_{2d}.
\end{equation}
This covariance matrix is clearly invertible, hence we can apply Theorem 1 to obtain:
\begin{equation}
    \begin{aligned}
    \Delta_{\rm C} \leq &C d^{\frac{1}{4}} \sum_{j=0}^{n-1} E(||(\frac{1}{2} I_{2n})^{-\frac{1}{2}} X_j^{(m)}||^3_2)\\
    = &2^{\frac{3}{2}} C d^{\frac{1}{4}} \sum_{j=0}^{n-1} E(||X_j^{(n)}||^3_2)\\
    = &2^{\frac{3}{2}} C d^{\frac{1}{4}} \sum_{j=0}^{n-1} \\
    &E(\sqrt{\frac{1}{n} \sum_{l=1}^d cos^2( \phi_j+2\pi\frac{j k_l}{n}) +sin^2(\phi_j+2\pi\frac{jk_l}{n})^3})\\
    = &2^{\frac{3}{2}} C d^{\frac{1}{4}} \sum_{j=0}^{n-1} E(\frac{n^{\frac{3}{2}}}{d^{\frac{3}{2}}}) = 2^{\frac{3}{2}} C d^{\frac{7}{4}}n^{-\frac{1}{2}}\\
    \leq &\acute{C} \frac{d^2}{\sqrt{n}}.
    \end{aligned}
\end{equation}
For a universal constant $\acute{C}>0$. In other words, the unitary constraint is upper bounded by $\frac{d^2}{\sqrt{n}}$. 

This result gives evidence that the number of modes must scale as $m\propto n^4$ to hide the unitary constraint. This shows the submatrix-of-circulant approaches the i.i.d. Gaussian ensemble additively. Unfortunately, unlike the case of Haar-random matrices, there appears to be no easy way to use the additive bound to prove the multiplicative bound.

\begin{figure*}
    \includegraphics[width=0.9\textwidth, keepaspectratio]{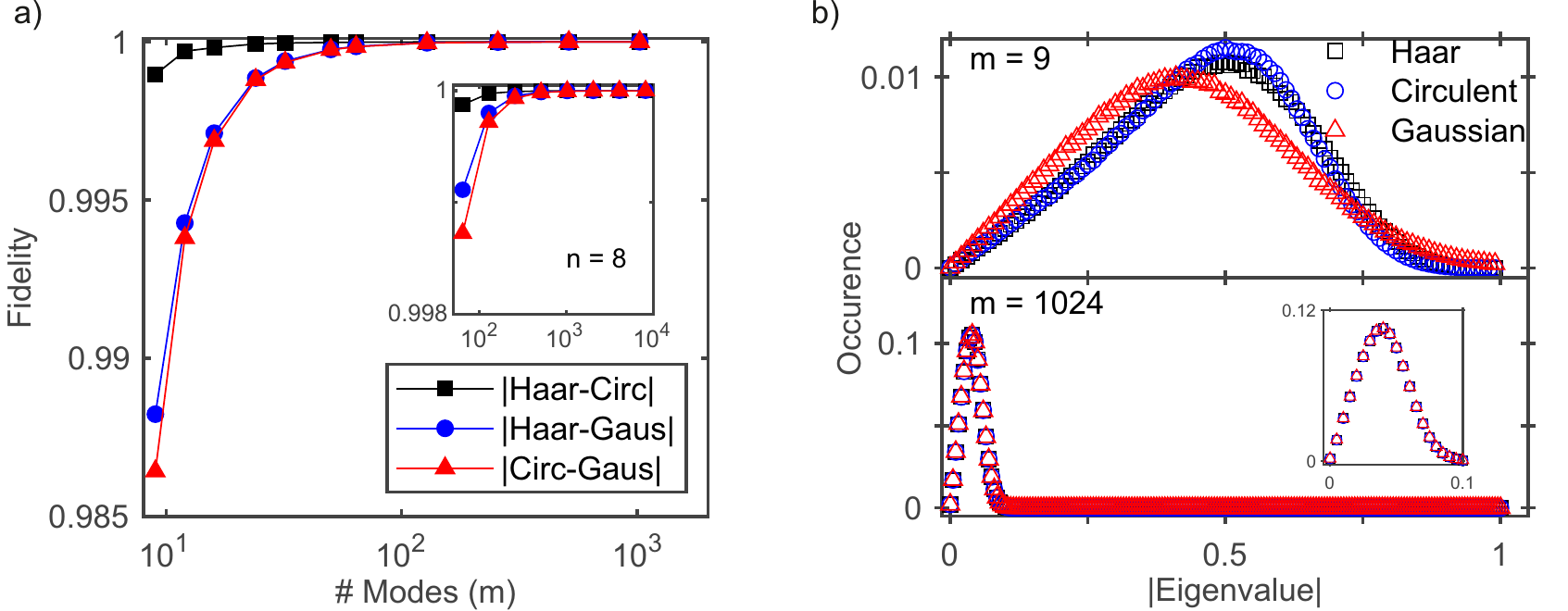}
    \caption{a) The convergence of the fidelity between eigenvalue distributions in the case of 3 photons. The inset shows the case for 8 photons. b) Two examples of the eigenvalue distribution for 3 photons when $m = 9$ and $m = 1024$. The inset on the bottom right is a zoom-in on the peak of the latter.}
    \label{fig:figEigExample}
\end{figure*}

\subsection{Eigenvalue spectra \label{ssecStrong}}
In the original boson sampling proposal, the multiplicative bound was found by examining the eigenvalue spectra.  \cite{mastrodonato_2007_LettMathPhys,petz_2005_Period.Math.Hung.,petz_2005_Probab.TheoryRelat.Fields,reffy_2005_undefined}. Inspired by this, we carry out a Monte-Carlo simulation of the eigenvalue spectra. To this end, $10^6$ truncated matrices of size $n$-by-$n$ are obtained, each from another Haar-random unitary and circulant unitary matrix of size $m$-by-$m$. Each truncation is chosen such that repeated entries are absent, i.e., it implicitly assumes an outcome without collisions. To quantify the similarity of the two distributions $A$ and $B$, we use the fidelity as given by: \cite{nielsen2010quantum}
\begin{equation}
    F( A, B) = \sum_{j=1}^n \sqrt{ P(A)_j P(B)_j} \ ,
\end{equation}
where $P(X)_j$ denotes the probability of the $j^{\rm th}$ eigenvalue. Note that this Monte-Carlo approach can only provide insights to the average fidelity between the eigenvalue spectra, it does not lower-bound the minimum fidelity. 

We start with a system of fixed photon number $n=3$, where we show how the fidelity of truncations of both Haar-random unitaries as well as circulant matrices with the i.i.d. Gaussian ensemble approach unity as the number of modes is increased. Figure \ref{fig:figEigExample}a) shows how the fidelity depends on the number of modes $m$ at a fixed photon number. Despite the differences between the truncated-Haar and truncated-circulant matrix ensembles, it is striking how similar their approach to $F = 1$ is as the number of modes is increased. That the distributions are indeed close is confirmed by the black points in Fig. \ref{fig:figEigExample}a), which show the fidelity between the Haar and circulant distributions, which show that they resemble each other more than that either resembles the i.i.d. Gaussian ensemble. The inset shows that this behavior remains when the number of photons is increased, in this case to $n=8$. The similarity of the fidelities as shown in \ref{fig:figEigExample}a) is illustrated by the two example eigenvalue distributions shown in Fig. \ref{fig:figEigExample}b). Again, the truncations of both the Haar-random and the circulant matrix are similar and in the limit of $m\gg n$ clearly converge to that of i.i.d. Gaussians.

\begin{figure}
    \includegraphics[width=0.45\textwidth, keepaspectratio]{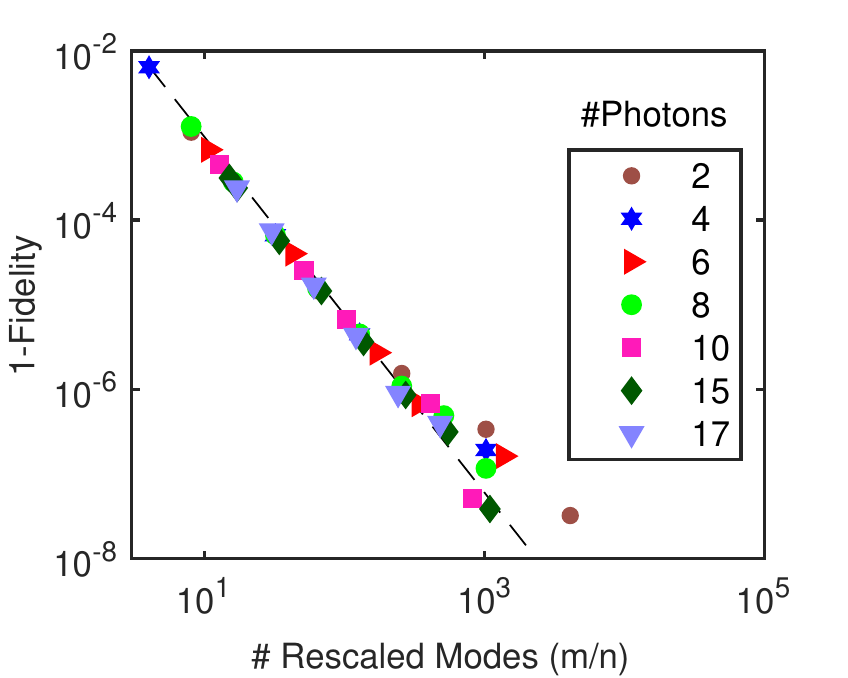}
    \caption{Convergence of the fidelity between circulant and Gaussian truncations with a rescaled x-axis. The dashed line is a guide to the eye.}
    \label{fig:figEigScaling}
\end{figure}

So far, we have considered eigenvalue distributions of truncated matrices at a fixed number of photons $n$ as a function of the number of modes. 
The next thing to show is how the ratio of modes and photons should scale for a fixed fidelity. This is shown in Fig. \ref{fig:figEigScaling}, where the convergence of the fidelity between the circulant and Gaussian truncation is plotted as a function of the number of modes $m$ for a different number of photons $n$. The x-axis is rescaled by the number of photons in order to emphasize the similar trend of the fidelity for all photons. The graph suggests that the number of modes should scale as $m \propto n$, however, note that this is the result after averaging over the space of circulant matrices; it is not the worst-case scenario. The deviation from the dashed line for a small number of photons is the result of using a finite number of samples in the simulation. This effect is more profound for small $n$ as each random matrix only contributes $n$ eigenvalues to the histogram.

\subsection{Permanents}
Another, more direct way to investigate the effect of the circulant correlations on the convergence to i.i.d. Gaussians is to look at the average outcome of permanents of submatrices of Haar-random and circulant matrices. If the correlations disappear with some suitable scaling of the number of modes and photons, the expected value of the permanent of a submatrix of a circulant $C$ must approach that of the i.i.d. Gaussian ensemble X, which has $E(|\mathrm{Perm}(X)|^2 = n!/{m^n}$. Having the average outcome of a permanent of a submatrix of a circulant matrix approach this value is a necessary but not a sufficient condition for the ensemble to approach that of i.i.d. Gaussians, but it can serve to provide numerical evidence.

\begin{figure}
    \includegraphics[width=0.45\textwidth, keepaspectratio]{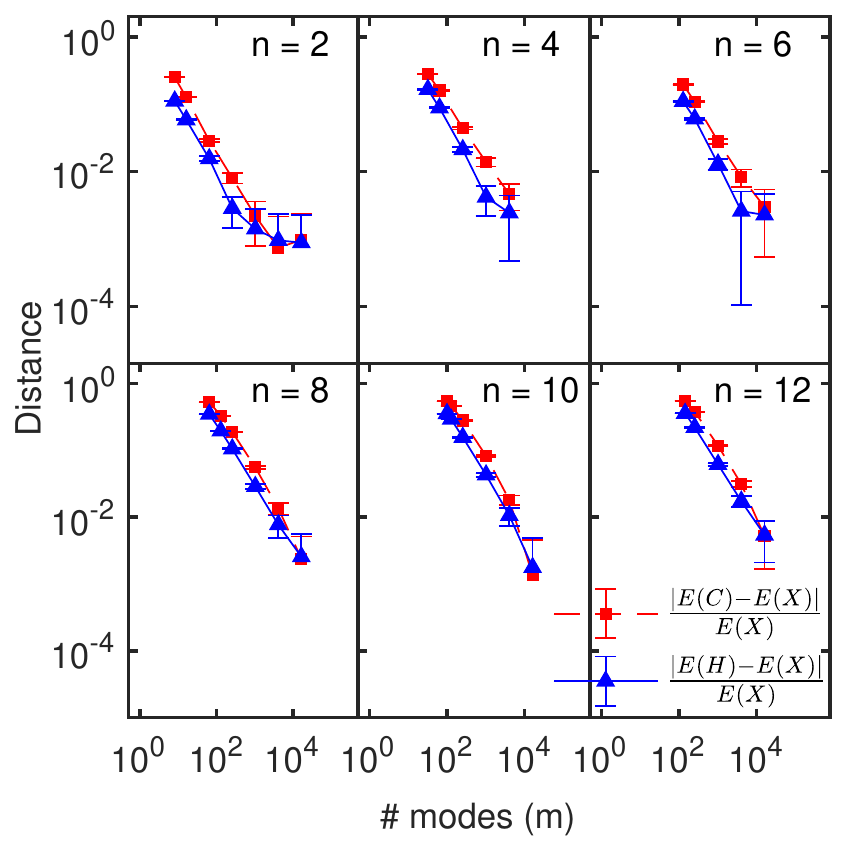}
    \caption{Numerical evidence for the convergence of the average permanent for a various number of photons $n$ as a function of the number of modes $m$. The red squares show how the distance between the average permanent of truncated circulant matrices and the result of i.i.d. Gaussian matrices diminishes with $m$ and the blue triangles show the same for truncated Haar matrices.}
    \label{fig:figPermanent}
\end{figure}

Figure \ref{fig:figPermanent} shows how the distance between the expectation values of truncations of both a Haar-random unitary and a circulant unitary with the result of an i.i.d. Gaussian scale with the number of modes, $m$. Each data point is the mean of $10^6$ random instances and the error bars denote the standard deviation. What is notable is that for each data point where we have sufficient numerical resolution in our simulation (as indicated by the error bars), the distance between Haar and Gaussian and between circulant and Gaussian differs only by a constant value as the number of modes is increased. 

This result reinforces the picture from our study of the eigenvalue spectrum, namely that the submatrix-of-circulant distribution strongly resembles the submatrix-of-Haar distribution, which is known to converge to i.i.d. Gaussian in the appropriate limits. A full proof would require a multiplicative version of the result of section \ref{ssecWeak}.

\section{No-collisions \label{secNoCol}}
The second prerequisite for our protocol is to show that the probability of observing collisions (i.e. outcomes with repeated matrix elements) becomes small with appropriate scaling of the number of modes $m$ versus the number of photons $n$. In the case of a Haar-random matrix, this can be done fairly straightforwardly, because each outcome, regardless of collisions, is equiprobable \cite{arkhipov_2012}. This reduces the problem to a matter of combinatorics, namely of counting the number of desired (no-collision) outcomes versus the number of total outcomes. 

\begin{figure}
    \begin{center}
    \includegraphics[width=0.45\textwidth, keepaspectratio]{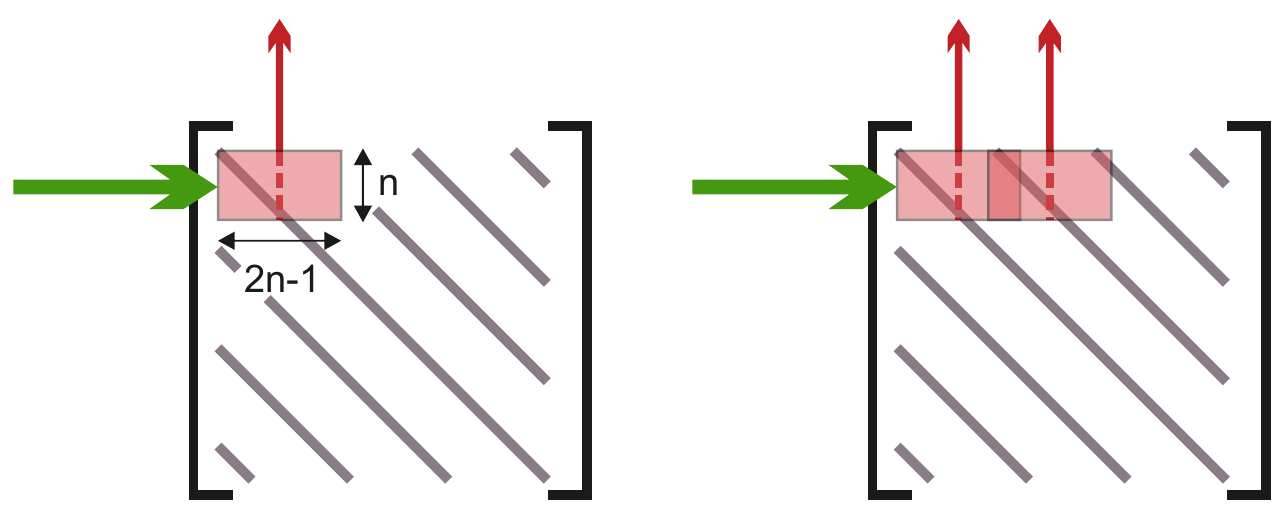} %
    \caption{Illustration of the counting argument where we assume an input (green) where photons enter in the first $n$ modes. If one of these photons comes out of, for example, output mode $n$ (red) then the other photons cannot get out of the first $2n-1$ output modes. (Left) The red block indicates what output modes are not allowed by $1$ photon leaving. (Right) The forbidden regions by two photons may overlap, as long as the used output modes (dashed lines) do not enter the forbidden region of the other photons.}
    \label{fig:circulantMatrixSampling_Abstract}
    \end{center}
\end{figure}

Unfortunately, for a circulant matrix, even in the limit where all elements are i.i.d. Gaussian, the outcomes are not on average equiprobable. The reason for this is that adjacent rows have repeated elements (see Fig. \ref{fig:circulantMatrixSampling_Abstract}). These repeated elements have two consequences: first, they have the effect that any outcome where two photons are detected $n$ modes apart will have repeated entries in its corresponding matrix $U_{\rm T}$, meaning that it will certainly not be i.i.d. Secondly, an elementary calculation shows that these repeated elements cause higher-order moments of Gaussians to appear in the calculation, which affect the expected value, making some outcomes more likely than others. 

To investigate the scaling of the no-collision condition with the number of photons and the number of modes, we follow the intuition of the previous section and make use of the fact that we have observed that the permanents of submatrices of circulant matrices without repeated elements resemble submatrices of Haar-random matrices in their behavior. We will start with a simple counting argument to find an analytical expression for the fraction of desired outcomes. Then we follow this up with numerical simulations to show that the intuition derived from this argument is correct.

\subsection{Number of good outcomes \label{subsec:Ngood}}
The number of desired outcomes, i.e., without circulant correlations, can be found by a counting argument. For now, we assume for simplicity that all outcomes are equiprobable. To show this counting argument, assume without loss of generality that the photons are incident on the first $n$ modes of the network. Each photon blocks exactly $2n-1$ output modes: on both sides of the output mode the first $n-1$ adjacent modes are not allowed since they contain the same value in the matrix and another $+1$ because a photon also blocks the its own output mode. 
See Fig. \ref{fig:circulantMatrixSampling_Abstract} for a schematic representation. These blocked regions may overlap, as long as none of the output modes are in these regions. Hence, in this limit, the fraction of good outcomes is given by:
\begin{equation}
    N_{\rm{good}} = \frac{m!_{2n-1}/(m-2n^2+n_{2n-1})}{\binom{m+n-1}{n}},\\
\end{equation}
where the subscripts denotes the $x^{th}$ multifactorial, i.e. $n!_{x} = n(n-x)(n-2x)(n-3x)...$. Working out these terms and only taking the first two leading terms ($n^k$ and $n^{k-1}$) in both the numerator and denominator results in:
\begin{equation}
    \begin{aligned}
    N_{\rm{good}} &= \frac{ m(m-2n+1)(m-4n+2)(m-6n+2)...}{(m+n-1)(m+n-2)(m+n-3)...}\\
    &\approx \frac{m^n -m^{n-1}\sum_{j=0}^{n-1} j(2n-1)}{m^n+m^{n-1}\sum_{j=0}^{n-1}{(n-j)}}\\
    &\approx \frac{m^n -\frac{1}{2}(n-1)(n-2)(2n-1)m^{n-1}}{m^n +\frac{1}{2}(n+1)n m^{n-1}}\\
    &\approx \frac{m^n -n^3 m^{n-1}}{m^n+n^2 m^{n-1}}.\\
    \end{aligned}
\end{equation}
Here the standard series $\sum_{j=0}^{j=n} n = n\frac{n-1}{2}$ has been used. As we are interested in the regime $m\gg n$, we can rewrite it as a Laurent series at $m=\infty$:
\begin{equation}
    \begin{aligned}
    N_{\rm{good}} &\approx \frac{1 -n^3 /m}{1+n^2 /m}\\
    &= 1 - \sum_{j=1}^{\infty} (-1)^j \frac{n^{2j}(n+1)}{m^j}\\
    &\approx 1 - \frac{n^3+n^2}{m} + \mathcal{O}(\frac{n^5}{m^2}).
    \end{aligned}
    \label{eq:NgoodScaling}
\end{equation}
This shows that the number of modes required to make the number of good (uncorrelated) outcomes dominate the experiment scales as $m \propto n^3$. The $n^3$ scaling is only a small increase from the normal $m \propto n^2$ scaling to hide the collisions. This is remarkable because there are additional correlations present in the circulant matrix and that a circulant matrix only has $n$, not $n^2$ unique elements.

\subsection{Numerical results}
\begin{figure*}
    \centering
    \includegraphics{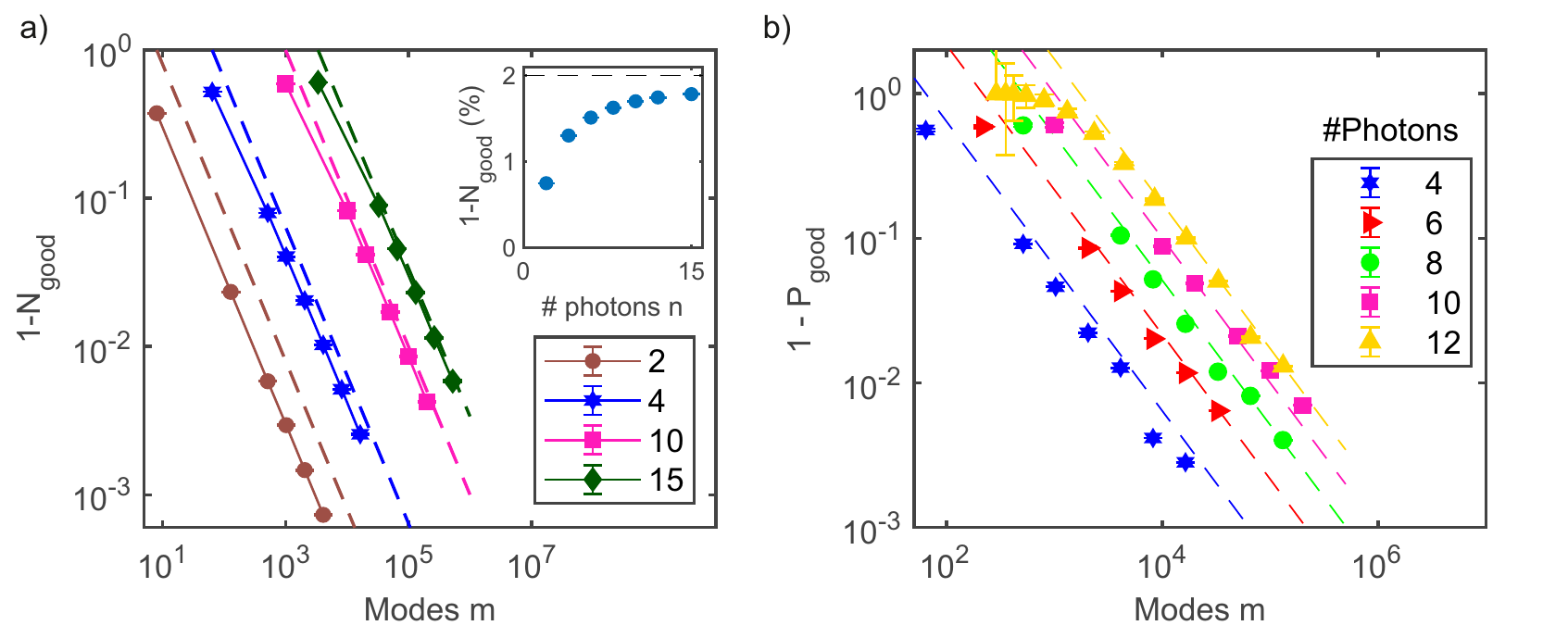}
    \caption{a) Monte-Carlo calculation of good outcomes for a selected number of photons. The dashed line is the predicted scaling. Note that the numerical results approach the theory for a large number of photons. Inset: the convergence to the fraction predicted by the combinatorics (dashed line) at $m/n^3=50$.
    b) Monte-Carlo calculation of the number of good outcomes, multiplied with the average probability of a good outcome. The error bars denote the standard deviation of $P_{\rm{good}}$. The dashed lines are guides to the eye showing the $\propto n^3 m^{-1}$ scaling.}
    \label{fig:NgoodPgood}
\end{figure*}

The above derived scaling for the fraction of possible outcomes without circulant correlations can be verified numerically. Figure \ref{fig:NgoodPgood}a) shows a Monte-Carlo simulation on how the fraction of correlated outcomes becomes small for some selected number of photons. Each data point is the average of $5\cdot 10^6$ samples, all drawn from independent instances of circulant unitaries. The error bars denote the standard deviation corresponding to a Bernoulli distribution. Each data set has a dashed line indicating the predicted $\frac{1}{n^3}$ scaling. The figure hints at a $m \propto n^3$ scaling, where the deviation for small photon numbers can be understood as finite size effects and the fact that no uncorrelated output states are possible when $m = n^2$. These results confirm the scaling as predicted by the combinatorics of section \ref{subsec:Ngood}.

The next step to show no-collisions is to identify the probability of ending up with an uncorrelated outcome and to show that this approaches 1 by increasing the total network size $m$. Therefore we first identify the number of uncorrelated (good) outcomes, just as in Fig. \ref{fig:NgoodPgood}a, and then multiply that with the average probability associated with these good outcomes. 

Figure \ref{fig:NgoodPgood}b) shows how the probability of obtaining a correlated outcome scales with the number of photons and the number of modes. Each data point is the product of the fraction of good outcomes, similar to Fig. \ref{fig:NgoodPgood}a), and the corresponding average probability of such outcome. Each sample is drawn from a circulant unitary and hence incorporates the unitary constraint. Note that this is a more stringent scenario than derived above in section \ref{subsec:Ngood}. The error bars denote the standard deviation, where again a Bernoulli distribution is used to estimate the uncertainty in $N_{\rm good}$. For each dataset -number of photons-, there is a dashed line indicating the predicted $m \propto n^3$ scaling.

\section{Required hardware}\label{secHardware}
In this section, we comment on the availability of the hardware needed to implement our proposal. 

A key experimental advantage of our proposal is that it removes the need for fiber-coupled single-photon sources. For example, in photon sources based on parametric down-conversion, a key source of optical loss is the coupling from the photons into an optical fiber. In our protocol, the photon sources could directly illuminate the scatterer, removing the need for fiber coupling and its associated optical loss.

To implement the random phases, there are three natural candidates: ground glass plates, holographic plates, or spatial light modulators. The first candidate for implementing such random phases, would be a surface scatterer such as a ground glass plate. This naturally implements the phases by random height differences on the surface, which corresponds to a relative phase delay. The second method is to use a holographic plate, made of glass or polymer where the thickness varies throughout the plate. By varying the thickness of the object locally, the light picks up a relative phase depending on the thickness. This method also gives a fixed phase shift, but allows for some degree of design. The transmission of a ground glass plate or a holographic plate is limited by the impedance mismatch between the material used and air, resulting in reflection losses of a few percent of the light. The third method is a fully tunable version of a phase-plate: a phase-only spatial light modulator (SLM). A phase-only SLM typically consists of an LCD screen in which every pixel can give a variable delay to the incident light on that pixel, implementing a phase shift. As SLMs are versatile and induce low losses -up to $90\%$ transmission efficiency-, SLMs are widely used in scattering adaptive-optics research \cite{mosk_2012_NatPhoton}. 

Once the photons went through the optical network, they must be detected. For this, a single-photon sensitive camera is needed as the multi-photon interference manifests itself in the transverse plane. Since lumping different output modes together will result in decoherence of the interference pattern, the total number of modes from in the network must be smaller or equal to that of the detector, and each detector pixel should correspond with a diffraction-limited spot. Hence circulant boson sampling requires megapixel single-photon cameras, which is the major technological limitation of our protocol. 

Superconducting nanowire single-photon detectors (SNSPD)s are a promising technology for such detectors. SNSPDs are reported to have record-high detection efficiencies \cite{lita_2008_OptExpress,reddy_2019} for a single detection pixel. Currently SNSPD based cameras are being developed and up to $64$ pixels are already reported on a regular basis \cite{doerner_2017_Appl.Phys.Lett.,gaggero_2019_Optica,miki_2014_Opt.ExpressOE,miyajima_2018_Opt.ExpressOE,sinclair_2019_IEEETrans.Appl.Supercond.,zhao_2013_Appl.Phys.Lett.,zhao_2017_Nat.Photonics,zhu_2018_Nat.Nanotechnol.,allmaras_2020_NanoLett.}. The first kilopixel detector is already reported as well \cite{wollman_2019_Opt.ExpressOE}, and current electronics architecture is known to allow up to $225\times 225$ pixels \cite{allman_2015_Appl.Phys.Lett.}. The detection efficiency of these devices is limited by the low fill factor on the sensor, typically $\approx 30\%$ and relative low quantum efficiency, which is typically in the order of $\approx 30-40\%$. It is estimated that the fill factor, however, can be improved to $\approx 90\%$ with already existing technology \cite{wollman_2019_Opt.ExpressOE}. Furthermore, the detection efficiency can be improved greatly by also placing each of the detection pixels inside an optical cavity. Note that this is the default method to boost detection efficiencies close to 100\% in single-pixel SNSPDs \cite{reddy_2019}.

\section{Conclusion}
In conclusion, we have described and theoretically analysed a free-space optical method for implementing random matrices suitable for boson sampling in a way that induces only minimal loss while making many optical modes accessible. We anticipate that this method will be useful for high-optical transmission quantum advantage demonstrations in photonics.

\begin{acknowledgments}
RvdM and PWHP acknowledges funding from the Duthc Research Council (NWO) via QuantERA QUOMPLEX (Grant No. 680.91.037), and the Dutch Research Counil (NWO), Veni programme 'Photonic Quantum Simulation', grant nr. 016.Veni.192.121.
 Furthermore we thank Peter Drummond for discussions.
\end{acknowledgments}

\bibliography{references}

\begin{thebibliography}{64}%
\makeatletter
\providecommand \@ifxundefined [1]{%
 \@ifx{#1\undefined}
}%
\providecommand \@ifnum [1]{%
 \ifnum #1\expandafter \@firstoftwo
 \else \expandafter \@secondoftwo
 \fi
}%
\providecommand \@ifx [1]{%
 \ifx #1\expandafter \@firstoftwo
 \else \expandafter \@secondoftwo
 \fi
}%
\providecommand \natexlab [1]{#1}%
\providecommand \enquote  [1]{``#1''}%
\providecommand \bibnamefont  [1]{#1}%
\providecommand \bibfnamefont [1]{#1}%
\providecommand \citenamefont [1]{#1}%
\providecommand \href@noop [0]{\@secondoftwo}%
\providecommand \href [0]{\begingroup \@sanitize@url \@href}%
\providecommand \@href[1]{\@@startlink{#1}\@@href}%
\providecommand \@@href[1]{\endgroup#1\@@endlink}%
\providecommand \@sanitize@url [0]{\catcode `\\12\catcode `\$12\catcode
  `\&12\catcode `\#12\catcode `\^12\catcode `\_12\catcode `\%12\relax}%
\providecommand \@@startlink[1]{}%
\providecommand \@@endlink[0]{}%
\providecommand \url  [0]{\begingroup\@sanitize@url \@url }%
\providecommand \@url [1]{\endgroup\@href {#1}{\urlprefix }}%
\providecommand \urlprefix  [0]{URL }%
\providecommand \Eprint [0]{\href }%
\providecommand \doibase [0]{http://dx.doi.org/}%
\providecommand \selectlanguage [0]{\@gobble}%
\providecommand \bibinfo  [0]{\@secondoftwo}%
\providecommand \bibfield  [0]{\@secondoftwo}%
\providecommand \translation [1]{[#1]}%
\providecommand \BibitemOpen [0]{}%
\providecommand \bibitemStop [0]{}%
\providecommand \bibitemNoStop [0]{.\EOS\space}%
\providecommand \EOS [0]{\spacefactor3000\relax}%
\providecommand \BibitemShut  [1]{\csname bibitem#1\endcsname}%
\let\auto@bib@innerbib\@empty
\bibitem [{\citenamefont {Arute}\ \emph {et~al.}(2019)\citenamefont {Arute},
  \citenamefont {Arya}, \citenamefont {Babbush}, \citenamefont {Bacon},
  \citenamefont {Bardin}, \citenamefont {Barends}, \citenamefont {Biswas},
  \citenamefont {Boixo}, \citenamefont {Brandao}, \citenamefont {Buell},
  \citenamefont {Burkett}, \citenamefont {Chen}, \citenamefont {Chen},
  \citenamefont {Chiaro}, \citenamefont {Collins}, \citenamefont {Courtney},
  \citenamefont {Dunsworth}, \citenamefont {Farhi}, \citenamefont {Foxen},
  \citenamefont {Fowler}, \citenamefont {Gidney}, \citenamefont {Giustina},
  \citenamefont {Graff}, \citenamefont {Guerin}, \citenamefont {Habegger},
  \citenamefont {Harrigan}, \citenamefont {Hartmann}, \citenamefont {Ho},
  \citenamefont {Hoffmann}, \citenamefont {Huang}, \citenamefont {Humble},
  \citenamefont {Isakov}, \citenamefont {Jeffrey}, \citenamefont {Jiang},
  \citenamefont {Kafri}, \citenamefont {Kechedzhi}, \citenamefont {Kelly},
  \citenamefont {Klimov}, \citenamefont {Knysh}, \citenamefont {Korotkov},
  \citenamefont {Kostritsa}, \citenamefont {Landhuis}, \citenamefont
  {Lindmark}, \citenamefont {Lucero}, \citenamefont {Lyakh}, \citenamefont
  {Mandr{\`a}}, \citenamefont {McClean}, \citenamefont {McEwen}, \citenamefont
  {Megrant}, \citenamefont {Mi}, \citenamefont {Michielsen}, \citenamefont
  {Mohseni}, \citenamefont {Mutus}, \citenamefont {Naaman}, \citenamefont
  {Neeley}, \citenamefont {Neill}, \citenamefont {Niu}, \citenamefont {Ostby},
  \citenamefont {Petukhov}, \citenamefont {Platt}, \citenamefont {Quintana},
  \citenamefont {Rieffel}, \citenamefont {Roushan}, \citenamefont {Rubin},
  \citenamefont {Sank}, \citenamefont {Satzinger}, \citenamefont {Smelyanskiy},
  \citenamefont {Sung}, \citenamefont {Trevithick}, \citenamefont
  {Vainsencher}, \citenamefont {Villalonga}, \citenamefont {White},
  \citenamefont {Yao}, \citenamefont {Yeh}, \citenamefont {Zalcman},
  \citenamefont {Neven},\ and\ \citenamefont {Martinis}}]{arute_2019_Nature}%
  \BibitemOpen
  \bibfield  {author} {\bibinfo {author} {\bibfnamefont {F.}~\bibnamefont
  {Arute}}, \bibinfo {author} {\bibfnamefont {K.}~\bibnamefont {Arya}},
  \bibinfo {author} {\bibfnamefont {R.}~\bibnamefont {Babbush}}, \bibinfo
  {author} {\bibfnamefont {D.}~\bibnamefont {Bacon}}, \bibinfo {author}
  {\bibfnamefont {J.~C.}\ \bibnamefont {Bardin}}, \bibinfo {author}
  {\bibfnamefont {R.}~\bibnamefont {Barends}}, \bibinfo {author} {\bibfnamefont
  {R.}~\bibnamefont {Biswas}}, \bibinfo {author} {\bibfnamefont
  {S.}~\bibnamefont {Boixo}}, \bibinfo {author} {\bibfnamefont {F.~G. S.~L.}\
  \bibnamefont {Brandao}}, \bibinfo {author} {\bibfnamefont {D.~A.}\
  \bibnamefont {Buell}}, \bibinfo {author} {\bibfnamefont {B.}~\bibnamefont
  {Burkett}}, \bibinfo {author} {\bibfnamefont {Y.}~\bibnamefont {Chen}},
  \bibinfo {author} {\bibfnamefont {Z.}~\bibnamefont {Chen}}, \bibinfo {author}
  {\bibfnamefont {B.}~\bibnamefont {Chiaro}}, \bibinfo {author} {\bibfnamefont
  {R.}~\bibnamefont {Collins}}, \bibinfo {author} {\bibfnamefont
  {W.}~\bibnamefont {Courtney}}, \bibinfo {author} {\bibfnamefont
  {A.}~\bibnamefont {Dunsworth}}, \bibinfo {author} {\bibfnamefont
  {E.}~\bibnamefont {Farhi}}, \bibinfo {author} {\bibfnamefont
  {B.}~\bibnamefont {Foxen}}, \bibinfo {author} {\bibfnamefont
  {A.}~\bibnamefont {Fowler}}, \bibinfo {author} {\bibfnamefont
  {C.}~\bibnamefont {Gidney}}, \bibinfo {author} {\bibfnamefont
  {M.}~\bibnamefont {Giustina}}, \bibinfo {author} {\bibfnamefont
  {R.}~\bibnamefont {Graff}}, \bibinfo {author} {\bibfnamefont
  {K.}~\bibnamefont {Guerin}}, \bibinfo {author} {\bibfnamefont
  {S.}~\bibnamefont {Habegger}}, \bibinfo {author} {\bibfnamefont {M.~P.}\
  \bibnamefont {Harrigan}}, \bibinfo {author} {\bibfnamefont {M.~J.}\
  \bibnamefont {Hartmann}}, \bibinfo {author} {\bibfnamefont {A.}~\bibnamefont
  {Ho}}, \bibinfo {author} {\bibfnamefont {M.}~\bibnamefont {Hoffmann}},
  \bibinfo {author} {\bibfnamefont {T.}~\bibnamefont {Huang}}, \bibinfo
  {author} {\bibfnamefont {T.~S.}\ \bibnamefont {Humble}}, \bibinfo {author}
  {\bibfnamefont {S.~V.}\ \bibnamefont {Isakov}}, \bibinfo {author}
  {\bibfnamefont {E.}~\bibnamefont {Jeffrey}}, \bibinfo {author} {\bibfnamefont
  {Z.}~\bibnamefont {Jiang}}, \bibinfo {author} {\bibfnamefont
  {D.}~\bibnamefont {Kafri}}, \bibinfo {author} {\bibfnamefont
  {K.}~\bibnamefont {Kechedzhi}}, \bibinfo {author} {\bibfnamefont
  {J.}~\bibnamefont {Kelly}}, \bibinfo {author} {\bibfnamefont {P.~V.}\
  \bibnamefont {Klimov}}, \bibinfo {author} {\bibfnamefont {S.}~\bibnamefont
  {Knysh}}, \bibinfo {author} {\bibfnamefont {A.}~\bibnamefont {Korotkov}},
  \bibinfo {author} {\bibfnamefont {F.}~\bibnamefont {Kostritsa}}, \bibinfo
  {author} {\bibfnamefont {D.}~\bibnamefont {Landhuis}}, \bibinfo {author}
  {\bibfnamefont {M.}~\bibnamefont {Lindmark}}, \bibinfo {author}
  {\bibfnamefont {E.}~\bibnamefont {Lucero}}, \bibinfo {author} {\bibfnamefont
  {D.}~\bibnamefont {Lyakh}}, \bibinfo {author} {\bibfnamefont
  {S.}~\bibnamefont {Mandr{\`a}}}, \bibinfo {author} {\bibfnamefont {J.~R.}\
  \bibnamefont {McClean}}, \bibinfo {author} {\bibfnamefont {M.}~\bibnamefont
  {McEwen}}, \bibinfo {author} {\bibfnamefont {A.}~\bibnamefont {Megrant}},
  \bibinfo {author} {\bibfnamefont {X.}~\bibnamefont {Mi}}, \bibinfo {author}
  {\bibfnamefont {K.}~\bibnamefont {Michielsen}}, \bibinfo {author}
  {\bibfnamefont {M.}~\bibnamefont {Mohseni}}, \bibinfo {author} {\bibfnamefont
  {J.}~\bibnamefont {Mutus}}, \bibinfo {author} {\bibfnamefont
  {O.}~\bibnamefont {Naaman}}, \bibinfo {author} {\bibfnamefont
  {M.}~\bibnamefont {Neeley}}, \bibinfo {author} {\bibfnamefont
  {C.}~\bibnamefont {Neill}}, \bibinfo {author} {\bibfnamefont {M.~Y.}\
  \bibnamefont {Niu}}, \bibinfo {author} {\bibfnamefont {E.}~\bibnamefont
  {Ostby}}, \bibinfo {author} {\bibfnamefont {A.}~\bibnamefont {Petukhov}},
  \bibinfo {author} {\bibfnamefont {J.~C.}\ \bibnamefont {Platt}}, \bibinfo
  {author} {\bibfnamefont {C.}~\bibnamefont {Quintana}}, \bibinfo {author}
  {\bibfnamefont {E.~G.}\ \bibnamefont {Rieffel}}, \bibinfo {author}
  {\bibfnamefont {P.}~\bibnamefont {Roushan}}, \bibinfo {author} {\bibfnamefont
  {N.~C.}\ \bibnamefont {Rubin}}, \bibinfo {author} {\bibfnamefont
  {D.}~\bibnamefont {Sank}}, \bibinfo {author} {\bibfnamefont {K.~J.}\
  \bibnamefont {Satzinger}}, \bibinfo {author} {\bibfnamefont {V.}~\bibnamefont
  {Smelyanskiy}}, \bibinfo {author} {\bibfnamefont {K.~J.}\ \bibnamefont
  {Sung}}, \bibinfo {author} {\bibfnamefont {M.~D.}\ \bibnamefont
  {Trevithick}}, \bibinfo {author} {\bibfnamefont {A.}~\bibnamefont
  {Vainsencher}}, \bibinfo {author} {\bibfnamefont {B.}~\bibnamefont
  {Villalonga}}, \bibinfo {author} {\bibfnamefont {T.}~\bibnamefont {White}},
  \bibinfo {author} {\bibfnamefont {Z.~J.}\ \bibnamefont {Yao}}, \bibinfo
  {author} {\bibfnamefont {P.}~\bibnamefont {Yeh}}, \bibinfo {author}
  {\bibfnamefont {A.}~\bibnamefont {Zalcman}}, \bibinfo {author} {\bibfnamefont
  {H.}~\bibnamefont {Neven}}, \ and\ \bibinfo {author} {\bibfnamefont {J.~M.}\
  \bibnamefont {Martinis}},\ }\href {\doibase 10.1038/s41586-019-1666-5}
  {\bibfield  {journal} {\bibinfo  {journal} {Nature}\ ,\ \bibinfo {pages} {1}}
  (\bibinfo {year} {2019})}\BibitemShut {NoStop}%
\bibitem [{\citenamefont {Zhong}\ \emph {et~al.}(2020)\citenamefont {Zhong},
  \citenamefont {Wang}, \citenamefont {Deng}, \citenamefont {Chen},
  \citenamefont {Peng}, \citenamefont {Luo}, \citenamefont {Qin}, \citenamefont
  {Wu}, \citenamefont {Ding}, \citenamefont {Hu}, \citenamefont {Hu},
  \citenamefont {Yang}, \citenamefont {Zhang}, \citenamefont {Li},
  \citenamefont {Li}, \citenamefont {Jiang}, \citenamefont {Gan}, \citenamefont
  {Yang}, \citenamefont {You}, \citenamefont {Wang}, \citenamefont {Li},
  \citenamefont {Liu}, \citenamefont {Lu},\ and\ \citenamefont
  {Pan}}]{zhong_2020_Science}%
  \BibitemOpen
  \bibfield  {author} {\bibinfo {author} {\bibfnamefont {H.-S.}\ \bibnamefont
  {Zhong}}, \bibinfo {author} {\bibfnamefont {H.}~\bibnamefont {Wang}},
  \bibinfo {author} {\bibfnamefont {Y.-H.}\ \bibnamefont {Deng}}, \bibinfo
  {author} {\bibfnamefont {M.-C.}\ \bibnamefont {Chen}}, \bibinfo {author}
  {\bibfnamefont {L.-C.}\ \bibnamefont {Peng}}, \bibinfo {author}
  {\bibfnamefont {Y.-H.}\ \bibnamefont {Luo}}, \bibinfo {author} {\bibfnamefont
  {J.}~\bibnamefont {Qin}}, \bibinfo {author} {\bibfnamefont {D.}~\bibnamefont
  {Wu}}, \bibinfo {author} {\bibfnamefont {X.}~\bibnamefont {Ding}}, \bibinfo
  {author} {\bibfnamefont {Y.}~\bibnamefont {Hu}}, \bibinfo {author}
  {\bibfnamefont {P.}~\bibnamefont {Hu}}, \bibinfo {author} {\bibfnamefont
  {X.-Y.}\ \bibnamefont {Yang}}, \bibinfo {author} {\bibfnamefont {W.-J.}\
  \bibnamefont {Zhang}}, \bibinfo {author} {\bibfnamefont {H.}~\bibnamefont
  {Li}}, \bibinfo {author} {\bibfnamefont {Y.}~\bibnamefont {Li}}, \bibinfo
  {author} {\bibfnamefont {X.}~\bibnamefont {Jiang}}, \bibinfo {author}
  {\bibfnamefont {L.}~\bibnamefont {Gan}}, \bibinfo {author} {\bibfnamefont
  {G.}~\bibnamefont {Yang}}, \bibinfo {author} {\bibfnamefont {L.}~\bibnamefont
  {You}}, \bibinfo {author} {\bibfnamefont {Z.}~\bibnamefont {Wang}}, \bibinfo
  {author} {\bibfnamefont {L.}~\bibnamefont {Li}}, \bibinfo {author}
  {\bibfnamefont {N.-L.}\ \bibnamefont {Liu}}, \bibinfo {author} {\bibfnamefont
  {C.-Y.}\ \bibnamefont {Lu}}, \ and\ \bibinfo {author} {\bibfnamefont {J.-W.}\
  \bibnamefont {Pan}},\ }\href {\doibase 10.1126/science.abe8770} {\bibfield
  {journal} {\bibinfo  {journal} {Science}\ } (\bibinfo {year} {2020}),\
  10.1126/science.abe8770}\BibitemShut {NoStop}%
\bibitem [{\citenamefont {Zhong}\ \emph {et~al.}(2021)\citenamefont {Zhong},
  \citenamefont {Deng}, \citenamefont {Qin}, \citenamefont {Wang},
  \citenamefont {Chen}, \citenamefont {Peng}, \citenamefont {Luo},
  \citenamefont {Wu}, \citenamefont {Gong}, \citenamefont {Su}, \citenamefont
  {Hu}, \citenamefont {Hu}, \citenamefont {Yang}, \citenamefont {Zhang},
  \citenamefont {Li}, \citenamefont {Li}, \citenamefont {Jiang}, \citenamefont
  {Gan}, \citenamefont {Yang}, \citenamefont {You}, \citenamefont {Wang},
  \citenamefont {Li}, \citenamefont {Liu}, \citenamefont {Renema},
  \citenamefont {Lu},\ and\ \citenamefont {Pan}}]{zhong_2021_ArXiv}%
  \BibitemOpen
  \bibfield  {author} {\bibinfo {author} {\bibfnamefont {H.-S.}\ \bibnamefont
  {Zhong}}, \bibinfo {author} {\bibfnamefont {Y.-H.}\ \bibnamefont {Deng}},
  \bibinfo {author} {\bibfnamefont {J.}~\bibnamefont {Qin}}, \bibinfo {author}
  {\bibfnamefont {H.}~\bibnamefont {Wang}}, \bibinfo {author} {\bibfnamefont
  {M.-C.}\ \bibnamefont {Chen}}, \bibinfo {author} {\bibfnamefont {L.-C.}\
  \bibnamefont {Peng}}, \bibinfo {author} {\bibfnamefont {Y.-H.}\ \bibnamefont
  {Luo}}, \bibinfo {author} {\bibfnamefont {D.}~\bibnamefont {Wu}}, \bibinfo
  {author} {\bibfnamefont {S.-Q.}\ \bibnamefont {Gong}}, \bibinfo {author}
  {\bibfnamefont {H.}~\bibnamefont {Su}}, \bibinfo {author} {\bibfnamefont
  {Y.}~\bibnamefont {Hu}}, \bibinfo {author} {\bibfnamefont {P.}~\bibnamefont
  {Hu}}, \bibinfo {author} {\bibfnamefont {X.-Y.}\ \bibnamefont {Yang}},
  \bibinfo {author} {\bibfnamefont {W.-J.}\ \bibnamefont {Zhang}}, \bibinfo
  {author} {\bibfnamefont {H.}~\bibnamefont {Li}}, \bibinfo {author}
  {\bibfnamefont {Y.}~\bibnamefont {Li}}, \bibinfo {author} {\bibfnamefont
  {X.}~\bibnamefont {Jiang}}, \bibinfo {author} {\bibfnamefont
  {L.}~\bibnamefont {Gan}}, \bibinfo {author} {\bibfnamefont {G.}~\bibnamefont
  {Yang}}, \bibinfo {author} {\bibfnamefont {L.}~\bibnamefont {You}}, \bibinfo
  {author} {\bibfnamefont {Z.}~\bibnamefont {Wang}}, \bibinfo {author}
  {\bibfnamefont {L.}~\bibnamefont {Li}}, \bibinfo {author} {\bibfnamefont
  {N.-L.}\ \bibnamefont {Liu}}, \bibinfo {author} {\bibfnamefont
  {J.}~\bibnamefont {Renema}}, \bibinfo {author} {\bibfnamefont {C.-Y.}\
  \bibnamefont {Lu}}, \ and\ \bibinfo {author} {\bibfnamefont {J.-W.}\
  \bibnamefont {Pan}},\ }\href@noop {} {\bibfield  {journal} {\bibinfo
  {journal} {arXiv:2106.15534 [physics, physics:quant-ph]}\ } (\bibinfo {year}
  {2021})},\ \Eprint {http://arxiv.org/abs/2106.15534} {arXiv:2106.15534
  [physics, physics:quant-ph]} \BibitemShut {NoStop}%
\bibitem [{\citenamefont {Wu}\ \emph {et~al.}(2021)\citenamefont {Wu},
  \citenamefont {Bao}, \citenamefont {Cao}, \citenamefont {Chen}, \citenamefont
  {Chen}, \citenamefont {Chen}, \citenamefont {Chung}, \citenamefont {Deng},
  \citenamefont {Du}, \citenamefont {Fan}, \citenamefont {Gong}, \citenamefont
  {Guo}, \citenamefont {Guo}, \citenamefont {Guo}, \citenamefont {Han},
  \citenamefont {Hong}, \citenamefont {Huang}, \citenamefont {Huo},
  \citenamefont {Li}, \citenamefont {Li}, \citenamefont {Li}, \citenamefont
  {Li}, \citenamefont {Liang}, \citenamefont {Lin}, \citenamefont {Lin},
  \citenamefont {Qian}, \citenamefont {Qiao}, \citenamefont {Rong},
  \citenamefont {Su}, \citenamefont {Sun}, \citenamefont {Wang}, \citenamefont
  {Wang}, \citenamefont {Wu}, \citenamefont {Xu}, \citenamefont {Yan},
  \citenamefont {Yang}, \citenamefont {Yang}, \citenamefont {Ye}, \citenamefont
  {Yin}, \citenamefont {Ying}, \citenamefont {Yu}, \citenamefont {Zha},
  \citenamefont {Zhang}, \citenamefont {Zhang}, \citenamefont {Zhang},
  \citenamefont {Zhang}, \citenamefont {Zhao}, \citenamefont {Zhao},
  \citenamefont {Zhou}, \citenamefont {Zhu}, \citenamefont {Lu}, \citenamefont
  {Peng}, \citenamefont {Zhu},\ and\ \citenamefont {Pan}}]{wu_2021_ArXiv}%
  \BibitemOpen
  \bibfield  {author} {\bibinfo {author} {\bibfnamefont {Y.}~\bibnamefont
  {Wu}}, \bibinfo {author} {\bibfnamefont {W.-S.}\ \bibnamefont {Bao}},
  \bibinfo {author} {\bibfnamefont {S.}~\bibnamefont {Cao}}, \bibinfo {author}
  {\bibfnamefont {F.}~\bibnamefont {Chen}}, \bibinfo {author} {\bibfnamefont
  {M.-C.}\ \bibnamefont {Chen}}, \bibinfo {author} {\bibfnamefont
  {X.}~\bibnamefont {Chen}}, \bibinfo {author} {\bibfnamefont {T.-H.}\
  \bibnamefont {Chung}}, \bibinfo {author} {\bibfnamefont {H.}~\bibnamefont
  {Deng}}, \bibinfo {author} {\bibfnamefont {Y.}~\bibnamefont {Du}}, \bibinfo
  {author} {\bibfnamefont {D.}~\bibnamefont {Fan}}, \bibinfo {author}
  {\bibfnamefont {M.}~\bibnamefont {Gong}}, \bibinfo {author} {\bibfnamefont
  {C.}~\bibnamefont {Guo}}, \bibinfo {author} {\bibfnamefont {C.}~\bibnamefont
  {Guo}}, \bibinfo {author} {\bibfnamefont {S.}~\bibnamefont {Guo}}, \bibinfo
  {author} {\bibfnamefont {L.}~\bibnamefont {Han}}, \bibinfo {author}
  {\bibfnamefont {L.}~\bibnamefont {Hong}}, \bibinfo {author} {\bibfnamefont
  {H.-L.}\ \bibnamefont {Huang}}, \bibinfo {author} {\bibfnamefont {Y.-H.}\
  \bibnamefont {Huo}}, \bibinfo {author} {\bibfnamefont {L.}~\bibnamefont
  {Li}}, \bibinfo {author} {\bibfnamefont {N.}~\bibnamefont {Li}}, \bibinfo
  {author} {\bibfnamefont {S.}~\bibnamefont {Li}}, \bibinfo {author}
  {\bibfnamefont {Y.}~\bibnamefont {Li}}, \bibinfo {author} {\bibfnamefont
  {F.}~\bibnamefont {Liang}}, \bibinfo {author} {\bibfnamefont
  {C.}~\bibnamefont {Lin}}, \bibinfo {author} {\bibfnamefont {J.}~\bibnamefont
  {Lin}}, \bibinfo {author} {\bibfnamefont {H.}~\bibnamefont {Qian}}, \bibinfo
  {author} {\bibfnamefont {D.}~\bibnamefont {Qiao}}, \bibinfo {author}
  {\bibfnamefont {H.}~\bibnamefont {Rong}}, \bibinfo {author} {\bibfnamefont
  {H.}~\bibnamefont {Su}}, \bibinfo {author} {\bibfnamefont {L.}~\bibnamefont
  {Sun}}, \bibinfo {author} {\bibfnamefont {L.}~\bibnamefont {Wang}}, \bibinfo
  {author} {\bibfnamefont {S.}~\bibnamefont {Wang}}, \bibinfo {author}
  {\bibfnamefont {D.}~\bibnamefont {Wu}}, \bibinfo {author} {\bibfnamefont
  {Y.}~\bibnamefont {Xu}}, \bibinfo {author} {\bibfnamefont {K.}~\bibnamefont
  {Yan}}, \bibinfo {author} {\bibfnamefont {W.}~\bibnamefont {Yang}}, \bibinfo
  {author} {\bibfnamefont {Y.}~\bibnamefont {Yang}}, \bibinfo {author}
  {\bibfnamefont {Y.}~\bibnamefont {Ye}}, \bibinfo {author} {\bibfnamefont
  {J.}~\bibnamefont {Yin}}, \bibinfo {author} {\bibfnamefont {C.}~\bibnamefont
  {Ying}}, \bibinfo {author} {\bibfnamefont {J.}~\bibnamefont {Yu}}, \bibinfo
  {author} {\bibfnamefont {C.}~\bibnamefont {Zha}}, \bibinfo {author}
  {\bibfnamefont {C.}~\bibnamefont {Zhang}}, \bibinfo {author} {\bibfnamefont
  {H.}~\bibnamefont {Zhang}}, \bibinfo {author} {\bibfnamefont
  {K.}~\bibnamefont {Zhang}}, \bibinfo {author} {\bibfnamefont
  {Y.}~\bibnamefont {Zhang}}, \bibinfo {author} {\bibfnamefont
  {H.}~\bibnamefont {Zhao}}, \bibinfo {author} {\bibfnamefont {Y.}~\bibnamefont
  {Zhao}}, \bibinfo {author} {\bibfnamefont {L.}~\bibnamefont {Zhou}}, \bibinfo
  {author} {\bibfnamefont {Q.}~\bibnamefont {Zhu}}, \bibinfo {author}
  {\bibfnamefont {C.-Y.}\ \bibnamefont {Lu}}, \bibinfo {author} {\bibfnamefont
  {C.-Z.}\ \bibnamefont {Peng}}, \bibinfo {author} {\bibfnamefont
  {X.}~\bibnamefont {Zhu}}, \ and\ \bibinfo {author} {\bibfnamefont {J.-W.}\
  \bibnamefont {Pan}},\ }\href@noop {} {\bibfield  {journal} {\bibinfo
  {journal} {arXiv:2106.14734 [quant-ph]}\ } (\bibinfo {year} {2021})},\
  \Eprint {http://arxiv.org/abs/2106.14734} {arXiv:2106.14734 [quant-ph]}
  \BibitemShut {NoStop}%
\bibitem [{\citenamefont {Aaronson}\ and\ \citenamefont
  {Arkhipov}(2013)}]{aaronson_2013_TheoryComput}%
  \BibitemOpen
  \bibfield  {author} {\bibinfo {author} {\bibfnamefont {S.}~\bibnamefont
  {Aaronson}}\ and\ \bibinfo {author} {\bibfnamefont {A.}~\bibnamefont
  {Arkhipov}},\ }\href@noop {} {\bibfield  {journal} {\bibinfo  {journal}
  {Theory Comput.}\ }\textbf {\bibinfo {volume} {9}},\ \bibinfo {pages} {143}
  (\bibinfo {year} {2013})}\BibitemShut {NoStop}%
\bibitem [{\citenamefont {Ryser}(1963)}]{ryser_1963_}%
  \BibitemOpen
  \bibfield  {author} {\bibinfo {author} {\bibfnamefont {H.~J.}\ \bibnamefont
  {Ryser}},\ }\href@noop {} {\emph {\bibinfo {title} {Combinatorial
  {{Mathematics}} ({{Carus Mathematical Monographs No}}. 14)}}}\ (\bibinfo
  {publisher} {{Mathematical Assn of Amer}},\ \bibinfo {year}
  {1963})\BibitemShut {NoStop}%
\bibitem [{\citenamefont {Wu}\ \emph {et~al.}(2018)\citenamefont {Wu},
  \citenamefont {Liu}, \citenamefont {Zhang}, \citenamefont {Jin},
  \citenamefont {Wang}, \citenamefont {Wang},\ and\ \citenamefont
  {Yang}}]{wu_2018_Natl.Sci.Rev.}%
  \BibitemOpen
  \bibfield  {author} {\bibinfo {author} {\bibfnamefont {J.}~\bibnamefont
  {Wu}}, \bibinfo {author} {\bibfnamefont {Y.}~\bibnamefont {Liu}}, \bibinfo
  {author} {\bibfnamefont {B.}~\bibnamefont {Zhang}}, \bibinfo {author}
  {\bibfnamefont {X.}~\bibnamefont {Jin}}, \bibinfo {author} {\bibfnamefont
  {Y.}~\bibnamefont {Wang}}, \bibinfo {author} {\bibfnamefont {H.}~\bibnamefont
  {Wang}}, \ and\ \bibinfo {author} {\bibfnamefont {X.}~\bibnamefont {Yang}},\
  }\href {\doibase 10.1093/nsr/nwy079} {\bibfield  {journal} {\bibinfo
  {journal} {National Science Review}\ }\textbf {\bibinfo {volume} {5}},\
  \bibinfo {pages} {715} (\bibinfo {year} {2018})},\ \Eprint
  {http://arxiv.org/abs/1606.05836} {arXiv:1606.05836} \BibitemShut {NoStop}%
\bibitem [{\citenamefont {Spring}\ \emph {et~al.}(2013)\citenamefont {Spring},
  \citenamefont {Metcalf}, \citenamefont {Humphreys}, \citenamefont
  {Kolthammer}, \citenamefont {Jin}, \citenamefont {Barbieri}, \citenamefont
  {Datta}, \citenamefont {{Thomas-Peter}}, \citenamefont {Langford},
  \citenamefont {Kundys}, \citenamefont {Gates}, \citenamefont {Smith},
  \citenamefont {Smith},\ and\ \citenamefont {Walmsley}}]{spring_2013_Science}%
  \BibitemOpen
  \bibfield  {author} {\bibinfo {author} {\bibfnamefont {J.~B.}\ \bibnamefont
  {Spring}}, \bibinfo {author} {\bibfnamefont {B.~J.}\ \bibnamefont {Metcalf}},
  \bibinfo {author} {\bibfnamefont {P.~C.}\ \bibnamefont {Humphreys}}, \bibinfo
  {author} {\bibfnamefont {W.~S.}\ \bibnamefont {Kolthammer}}, \bibinfo
  {author} {\bibfnamefont {X.-M.}\ \bibnamefont {Jin}}, \bibinfo {author}
  {\bibfnamefont {M.}~\bibnamefont {Barbieri}}, \bibinfo {author}
  {\bibfnamefont {A.}~\bibnamefont {Datta}}, \bibinfo {author} {\bibfnamefont
  {N.}~\bibnamefont {{Thomas-Peter}}}, \bibinfo {author} {\bibfnamefont
  {N.~K.}\ \bibnamefont {Langford}}, \bibinfo {author} {\bibfnamefont
  {D.}~\bibnamefont {Kundys}}, \bibinfo {author} {\bibfnamefont {J.~C.}\
  \bibnamefont {Gates}}, \bibinfo {author} {\bibfnamefont {B.~J.}\ \bibnamefont
  {Smith}}, \bibinfo {author} {\bibfnamefont {P.~G.~R.}\ \bibnamefont {Smith}},
  \ and\ \bibinfo {author} {\bibfnamefont {I.~A.}\ \bibnamefont {Walmsley}},\
  }\href {\doibase 10.1126/science.1231692} {\bibfield  {journal} {\bibinfo
  {journal} {Science}\ }\textbf {\bibinfo {volume} {339}},\ \bibinfo {pages}
  {798} (\bibinfo {year} {2013})}\BibitemShut {NoStop}%
\bibitem [{\citenamefont {Broome}\ \emph {et~al.}(2013)\citenamefont {Broome},
  \citenamefont {Fedrizzi}, \citenamefont {{Rahimi-Keshari}}, \citenamefont
  {Dove}, \citenamefont {Aaronson}, \citenamefont {Ralph},\ and\ \citenamefont
  {White}}]{broome_2013_Science}%
  \BibitemOpen
  \bibfield  {author} {\bibinfo {author} {\bibfnamefont {M.~A.}\ \bibnamefont
  {Broome}}, \bibinfo {author} {\bibfnamefont {A.}~\bibnamefont {Fedrizzi}},
  \bibinfo {author} {\bibfnamefont {S.}~\bibnamefont {{Rahimi-Keshari}}},
  \bibinfo {author} {\bibfnamefont {J.}~\bibnamefont {Dove}}, \bibinfo {author}
  {\bibfnamefont {S.}~\bibnamefont {Aaronson}}, \bibinfo {author}
  {\bibfnamefont {T.~C.}\ \bibnamefont {Ralph}}, \ and\ \bibinfo {author}
  {\bibfnamefont {A.~G.}\ \bibnamefont {White}},\ }\href {\doibase
  10.1126/science.1231440} {\bibfield  {journal} {\bibinfo  {journal}
  {Science}\ }\textbf {\bibinfo {volume} {339}},\ \bibinfo {pages} {794}
  (\bibinfo {year} {2013})}\BibitemShut {NoStop}%
\bibitem [{\citenamefont {Tillmann}\ \emph {et~al.}(2013)\citenamefont
  {Tillmann}, \citenamefont {Daki{\'c}}, \citenamefont {Heilmann},
  \citenamefont {Nolte}, \citenamefont {Szameit},\ and\ \citenamefont
  {Walther}}]{tillmann_2013_NatPhotonics}%
  \BibitemOpen
  \bibfield  {author} {\bibinfo {author} {\bibfnamefont {M.}~\bibnamefont
  {Tillmann}}, \bibinfo {author} {\bibfnamefont {B.}~\bibnamefont {Daki{\'c}}},
  \bibinfo {author} {\bibfnamefont {R.}~\bibnamefont {Heilmann}}, \bibinfo
  {author} {\bibfnamefont {S.}~\bibnamefont {Nolte}}, \bibinfo {author}
  {\bibfnamefont {A.}~\bibnamefont {Szameit}}, \ and\ \bibinfo {author}
  {\bibfnamefont {P.}~\bibnamefont {Walther}},\ }\href@noop {} {\bibfield
  {journal} {\bibinfo  {journal} {Nat. Photonics}\ }\textbf {\bibinfo {volume}
  {7}},\ \bibinfo {pages} {540} (\bibinfo {year} {2013})}\BibitemShut {NoStop}%
\bibitem [{\citenamefont {Crespi}\ \emph {et~al.}(2013)\citenamefont {Crespi},
  \citenamefont {Osellame}, \citenamefont {Ramponi}, \citenamefont {Brod},
  \citenamefont {Galv{\~a}o}, \citenamefont {Spagnolo}, \citenamefont
  {Vitelli}, \citenamefont {Maiorino}, \citenamefont {Mataloni},\ and\
  \citenamefont {Sciarrino}}]{crespi_2013_NatPhotonics}%
  \BibitemOpen
  \bibfield  {author} {\bibinfo {author} {\bibfnamefont {A.}~\bibnamefont
  {Crespi}}, \bibinfo {author} {\bibfnamefont {R.}~\bibnamefont {Osellame}},
  \bibinfo {author} {\bibfnamefont {R.}~\bibnamefont {Ramponi}}, \bibinfo
  {author} {\bibfnamefont {D.~J.}\ \bibnamefont {Brod}}, \bibinfo {author}
  {\bibfnamefont {E.~F.}\ \bibnamefont {Galv{\~a}o}}, \bibinfo {author}
  {\bibfnamefont {N.}~\bibnamefont {Spagnolo}}, \bibinfo {author}
  {\bibfnamefont {C.}~\bibnamefont {Vitelli}}, \bibinfo {author} {\bibfnamefont
  {E.}~\bibnamefont {Maiorino}}, \bibinfo {author} {\bibfnamefont
  {P.}~\bibnamefont {Mataloni}}, \ and\ \bibinfo {author} {\bibfnamefont
  {F.}~\bibnamefont {Sciarrino}},\ }\href@noop {} {\bibfield  {journal}
  {\bibinfo  {journal} {Nat. Photonics}\ }\textbf {\bibinfo {volume} {7}},\
  \bibinfo {pages} {545} (\bibinfo {year} {2013})}\BibitemShut {NoStop}%
\bibitem [{\citenamefont {Loredo}\ \emph {et~al.}(2017)\citenamefont {Loredo},
  \citenamefont {Broome}, \citenamefont {Hilaire}, \citenamefont {Gazzano},
  \citenamefont {Sagnes}, \citenamefont {Lemaitre}, \citenamefont {Almeida},
  \citenamefont {Senellart},\ and\ \citenamefont {White}}]{loredo_2017_PRL}%
  \BibitemOpen
  \bibfield  {author} {\bibinfo {author} {\bibfnamefont {J.~C.}\ \bibnamefont
  {Loredo}}, \bibinfo {author} {\bibfnamefont {M.~A.}\ \bibnamefont {Broome}},
  \bibinfo {author} {\bibfnamefont {P.}~\bibnamefont {Hilaire}}, \bibinfo
  {author} {\bibfnamefont {O.}~\bibnamefont {Gazzano}}, \bibinfo {author}
  {\bibfnamefont {I.}~\bibnamefont {Sagnes}}, \bibinfo {author} {\bibfnamefont
  {A.}~\bibnamefont {Lemaitre}}, \bibinfo {author} {\bibfnamefont {M.~P.}\
  \bibnamefont {Almeida}}, \bibinfo {author} {\bibfnamefont {P.}~\bibnamefont
  {Senellart}}, \ and\ \bibinfo {author} {\bibfnamefont {A.~G.}\ \bibnamefont
  {White}},\ }\href {\doibase 10.1103/PhysRevLett.118.130503} {\bibfield
  {journal} {\bibinfo  {journal} {Phys. Rev. Lett.}\ }\textbf {\bibinfo
  {volume} {118}},\ \bibinfo {pages} {130503} (\bibinfo {year}
  {2017})}\BibitemShut {NoStop}%
\bibitem [{\citenamefont {Wang}\ \emph {et~al.}(2019)\citenamefont {Wang},
  \citenamefont {Qin}, \citenamefont {Ding}, \citenamefont {Chen},
  \citenamefont {Chen}, \citenamefont {You}, \citenamefont {He}, \citenamefont
  {Jiang}, \citenamefont {You}, \citenamefont {Wang}, \citenamefont
  {Schneider}, \citenamefont {Renema}, \citenamefont {H{\"o}fling},
  \citenamefont {Lu},\ and\ \citenamefont {Pan}}]{wang_2019_Phys.Rev.Lett.}%
  \BibitemOpen
  \bibfield  {author} {\bibinfo {author} {\bibfnamefont {H.}~\bibnamefont
  {Wang}}, \bibinfo {author} {\bibfnamefont {J.}~\bibnamefont {Qin}}, \bibinfo
  {author} {\bibfnamefont {X.}~\bibnamefont {Ding}}, \bibinfo {author}
  {\bibfnamefont {M.-C.}\ \bibnamefont {Chen}}, \bibinfo {author}
  {\bibfnamefont {S.}~\bibnamefont {Chen}}, \bibinfo {author} {\bibfnamefont
  {X.}~\bibnamefont {You}}, \bibinfo {author} {\bibfnamefont {Y.-M.}\
  \bibnamefont {He}}, \bibinfo {author} {\bibfnamefont {X.}~\bibnamefont
  {Jiang}}, \bibinfo {author} {\bibfnamefont {L.}~\bibnamefont {You}}, \bibinfo
  {author} {\bibfnamefont {Z.}~\bibnamefont {Wang}}, \bibinfo {author}
  {\bibfnamefont {C.}~\bibnamefont {Schneider}}, \bibinfo {author}
  {\bibfnamefont {J.~J.}\ \bibnamefont {Renema}}, \bibinfo {author}
  {\bibfnamefont {S.}~\bibnamefont {H{\"o}fling}}, \bibinfo {author}
  {\bibfnamefont {C.-Y.}\ \bibnamefont {Lu}}, \ and\ \bibinfo {author}
  {\bibfnamefont {J.-W.}\ \bibnamefont {Pan}},\ }\href {\doibase
  10.1103/PhysRevLett.123.250503} {\bibfield  {journal} {\bibinfo  {journal}
  {Phys. Rev. Lett.}\ }\textbf {\bibinfo {volume} {123}},\ \bibinfo {pages}
  {250503} (\bibinfo {year} {2019})}\BibitemShut {NoStop}%
\bibitem [{\citenamefont {He}\ \emph {et~al.}(2017)\citenamefont {He},
  \citenamefont {Ding}, \citenamefont {Su}, \citenamefont {Huang},
  \citenamefont {Qin}, \citenamefont {Wang}, \citenamefont {Unsleber},
  \citenamefont {Chen}, \citenamefont {Wang}, \citenamefont {He}, \citenamefont
  {Wang}, \citenamefont {Zhang}, \citenamefont {Chen}, \citenamefont
  {Schneider}, \citenamefont {Kamp}, \citenamefont {You}, \citenamefont {Wang},
  \citenamefont {H{\"o}fling}, \citenamefont {Lu},\ and\ \citenamefont
  {Pan}}]{he_2017_PhysRevLett}%
  \BibitemOpen
  \bibfield  {author} {\bibinfo {author} {\bibfnamefont {Y.}~\bibnamefont
  {He}}, \bibinfo {author} {\bibfnamefont {X.}~\bibnamefont {Ding}}, \bibinfo
  {author} {\bibfnamefont {Z.-E.}\ \bibnamefont {Su}}, \bibinfo {author}
  {\bibfnamefont {H.-L.}\ \bibnamefont {Huang}}, \bibinfo {author}
  {\bibfnamefont {J.}~\bibnamefont {Qin}}, \bibinfo {author} {\bibfnamefont
  {C.}~\bibnamefont {Wang}}, \bibinfo {author} {\bibfnamefont {S.}~\bibnamefont
  {Unsleber}}, \bibinfo {author} {\bibfnamefont {C.}~\bibnamefont {Chen}},
  \bibinfo {author} {\bibfnamefont {H.}~\bibnamefont {Wang}}, \bibinfo {author}
  {\bibfnamefont {Y.-M.}\ \bibnamefont {He}}, \bibinfo {author} {\bibfnamefont
  {X.-L.}\ \bibnamefont {Wang}}, \bibinfo {author} {\bibfnamefont {W.-J.}\
  \bibnamefont {Zhang}}, \bibinfo {author} {\bibfnamefont {S.-J.}\ \bibnamefont
  {Chen}}, \bibinfo {author} {\bibfnamefont {C.}~\bibnamefont {Schneider}},
  \bibinfo {author} {\bibfnamefont {M.}~\bibnamefont {Kamp}}, \bibinfo {author}
  {\bibfnamefont {L.-X.}\ \bibnamefont {You}}, \bibinfo {author} {\bibfnamefont
  {Z.}~\bibnamefont {Wang}}, \bibinfo {author} {\bibfnamefont {S.}~\bibnamefont
  {H{\"o}fling}}, \bibinfo {author} {\bibfnamefont {C.-Y.}\ \bibnamefont {Lu}},
  \ and\ \bibinfo {author} {\bibfnamefont {J.-W.}\ \bibnamefont {Pan}},\ }\href
  {\doibase 10.1103/PhysRevLett.118.190501} {\bibfield  {journal} {\bibinfo
  {journal} {Phys. Rev. Lett.}\ }\textbf {\bibinfo {volume} {118}},\ \bibinfo
  {pages} {190501} (\bibinfo {year} {2017})}\BibitemShut {NoStop}%
\bibitem [{\citenamefont {Zhong}\ \emph {et~al.}(2018)\citenamefont {Zhong},
  \citenamefont {Li}, \citenamefont {Li}, \citenamefont {Peng}, \citenamefont
  {Su}, \citenamefont {Hu}, \citenamefont {He}, \citenamefont {Ding},
  \citenamefont {Zhang}, \citenamefont {Li}, \citenamefont {Zhang},
  \citenamefont {Wang}, \citenamefont {You}, \citenamefont {Wang},
  \citenamefont {Jiang}, \citenamefont {Li}, \citenamefont {Chen},
  \citenamefont {Liu}, \citenamefont {Lu},\ and\ \citenamefont
  {Pan}}]{zhong_2018_PRL}%
  \BibitemOpen
  \bibfield  {author} {\bibinfo {author} {\bibfnamefont {H.-S.}\ \bibnamefont
  {Zhong}}, \bibinfo {author} {\bibfnamefont {Y.}~\bibnamefont {Li}}, \bibinfo
  {author} {\bibfnamefont {W.}~\bibnamefont {Li}}, \bibinfo {author}
  {\bibfnamefont {L.-C.}\ \bibnamefont {Peng}}, \bibinfo {author}
  {\bibfnamefont {Z.-E.}\ \bibnamefont {Su}}, \bibinfo {author} {\bibfnamefont
  {Y.}~\bibnamefont {Hu}}, \bibinfo {author} {\bibfnamefont {Y.-M.}\
  \bibnamefont {He}}, \bibinfo {author} {\bibfnamefont {X.}~\bibnamefont
  {Ding}}, \bibinfo {author} {\bibfnamefont {W.}~\bibnamefont {Zhang}},
  \bibinfo {author} {\bibfnamefont {H.}~\bibnamefont {Li}}, \bibinfo {author}
  {\bibfnamefont {L.}~\bibnamefont {Zhang}}, \bibinfo {author} {\bibfnamefont
  {Z.}~\bibnamefont {Wang}}, \bibinfo {author} {\bibfnamefont {L.}~\bibnamefont
  {You}}, \bibinfo {author} {\bibfnamefont {X.-L.}\ \bibnamefont {Wang}},
  \bibinfo {author} {\bibfnamefont {X.}~\bibnamefont {Jiang}}, \bibinfo
  {author} {\bibfnamefont {L.}~\bibnamefont {Li}}, \bibinfo {author}
  {\bibfnamefont {Y.-A.}\ \bibnamefont {Chen}}, \bibinfo {author}
  {\bibfnamefont {N.-L.}\ \bibnamefont {Liu}}, \bibinfo {author} {\bibfnamefont
  {C.-Y.}\ \bibnamefont {Lu}}, \ and\ \bibinfo {author} {\bibfnamefont {J.-W.}\
  \bibnamefont {Pan}},\ }\href {\doibase 10.1103/PhysRevLett.121.250505}
  {\bibfield  {journal} {\bibinfo  {journal} {Phys. Rev. Lett.}\ }\textbf
  {\bibinfo {volume} {121}},\ \bibinfo {pages} {250505} (\bibinfo {year}
  {2018})}\BibitemShut {NoStop}%
\bibitem [{\citenamefont {Lund}\ \emph {et~al.}(2014)\citenamefont {Lund},
  \citenamefont {Laing}, \citenamefont {{Rahimi-Keshari}}, \citenamefont
  {Rudolph}, \citenamefont {O'Brien},\ and\ \citenamefont
  {Ralph}}]{lund_2014_Phys.Rev.Lett.}%
  \BibitemOpen
  \bibfield  {author} {\bibinfo {author} {\bibfnamefont {A.~P.}\ \bibnamefont
  {Lund}}, \bibinfo {author} {\bibfnamefont {A.}~\bibnamefont {Laing}},
  \bibinfo {author} {\bibfnamefont {S.}~\bibnamefont {{Rahimi-Keshari}}},
  \bibinfo {author} {\bibfnamefont {T.}~\bibnamefont {Rudolph}}, \bibinfo
  {author} {\bibfnamefont {J.~L.}\ \bibnamefont {O'Brien}}, \ and\ \bibinfo
  {author} {\bibfnamefont {T.~C.}\ \bibnamefont {Ralph}},\ }\href {\doibase
  10.1103/PhysRevLett.113.100502} {\bibfield  {journal} {\bibinfo  {journal}
  {Phys. Rev. Lett.}\ }\textbf {\bibinfo {volume} {113}},\ \bibinfo {pages}
  {100502} (\bibinfo {year} {2014})}\BibitemShut {NoStop}%
\bibitem [{\citenamefont {Hamilton}\ \emph {et~al.}(2017)\citenamefont
  {Hamilton}, \citenamefont {Kruse}, \citenamefont {Sansoni}, \citenamefont
  {Barkhofen}, \citenamefont {Silberhorn},\ and\ \citenamefont
  {Jex}}]{hamilton_2017_Phys.Rev.Lett.}%
  \BibitemOpen
  \bibfield  {author} {\bibinfo {author} {\bibfnamefont {C.~S.}\ \bibnamefont
  {Hamilton}}, \bibinfo {author} {\bibfnamefont {R.}~\bibnamefont {Kruse}},
  \bibinfo {author} {\bibfnamefont {L.}~\bibnamefont {Sansoni}}, \bibinfo
  {author} {\bibfnamefont {S.}~\bibnamefont {Barkhofen}}, \bibinfo {author}
  {\bibfnamefont {C.}~\bibnamefont {Silberhorn}}, \ and\ \bibinfo {author}
  {\bibfnamefont {I.}~\bibnamefont {Jex}},\ }\href {\doibase
  10.1103/PhysRevLett.119.170501} {\bibfield  {journal} {\bibinfo  {journal}
  {Phys. Rev. Lett.}\ }\textbf {\bibinfo {volume} {119}},\ \bibinfo {pages}
  {170501} (\bibinfo {year} {2017})}\BibitemShut {NoStop}%
\bibitem [{\citenamefont {Renema}\ \emph
  {et~al.}(2018{\natexlab{a}})\citenamefont {Renema}, \citenamefont {Menssen},
  \citenamefont {Clements}, \citenamefont {Triginer}, \citenamefont
  {Kolthammer},\ and\ \citenamefont {Walmsley}}]{renema_2018_Phys.Rev.Lett.}%
  \BibitemOpen
  \bibfield  {author} {\bibinfo {author} {\bibfnamefont {J.~J.}\ \bibnamefont
  {Renema}}, \bibinfo {author} {\bibfnamefont {A.}~\bibnamefont {Menssen}},
  \bibinfo {author} {\bibfnamefont {W.~R.}\ \bibnamefont {Clements}}, \bibinfo
  {author} {\bibfnamefont {G.}~\bibnamefont {Triginer}}, \bibinfo {author}
  {\bibfnamefont {W.~S.}\ \bibnamefont {Kolthammer}}, \ and\ \bibinfo {author}
  {\bibfnamefont {I.~A.}\ \bibnamefont {Walmsley}},\ }\href {\doibase
  10.1103/PhysRevLett.120.220502} {\bibfield  {journal} {\bibinfo  {journal}
  {Phys. Rev. Lett.}\ }\textbf {\bibinfo {volume} {120}},\ \bibinfo {pages}
  {220502} (\bibinfo {year} {2018}{\natexlab{a}})}\BibitemShut {NoStop}%
\bibitem [{\citenamefont {Renema}\ \emph
  {et~al.}(2018{\natexlab{b}})\citenamefont {Renema}, \citenamefont
  {Shchesnovich},\ and\ \citenamefont {{Garcia-Patron}}}]{renema_2018_ArXiv}%
  \BibitemOpen
  \bibfield  {author} {\bibinfo {author} {\bibfnamefont {J.}~\bibnamefont
  {Renema}}, \bibinfo {author} {\bibfnamefont {V.}~\bibnamefont
  {Shchesnovich}}, \ and\ \bibinfo {author} {\bibfnamefont {R.}~\bibnamefont
  {{Garcia-Patron}}},\ }\href@noop {} {\bibfield  {journal} {\bibinfo
  {journal} {arXiv:1809.01953 [cond-mat, physics:physics, physics:quant-ph]}\ }
  (\bibinfo {year} {2018}{\natexlab{b}})},\ \Eprint
  {http://arxiv.org/abs/1809.01953} {arXiv:1809.01953 [cond-mat,
  physics:physics, physics:quant-ph]} \BibitemShut {NoStop}%
\bibitem [{\citenamefont {{Garc{\'i}a-Patr{\'o}n}}\ \emph
  {et~al.}(2019)\citenamefont {{Garc{\'i}a-Patr{\'o}n}}, \citenamefont
  {Renema},\ and\ \citenamefont {Shchesnovich}}]{garcia-patron_2019_Quantum}%
  \BibitemOpen
  \bibfield  {author} {\bibinfo {author} {\bibfnamefont {R.}~\bibnamefont
  {{Garc{\'i}a-Patr{\'o}n}}}, \bibinfo {author} {\bibfnamefont {J.~J.}\
  \bibnamefont {Renema}}, \ and\ \bibinfo {author} {\bibfnamefont
  {V.}~\bibnamefont {Shchesnovich}},\ }\href {\doibase
  10.22331/q-2019-08-05-169} {\bibfield  {journal} {\bibinfo  {journal}
  {Quantum}\ }\textbf {\bibinfo {volume} {3}},\ \bibinfo {pages} {169}
  (\bibinfo {year} {2019})},\ \Eprint {http://arxiv.org/abs/1712.10037}
  {arXiv:1712.10037} \BibitemShut {NoStop}%
\bibitem [{\citenamefont {Brod}\ and\ \citenamefont
  {Oszmaniec}(2020)}]{brod_2020_Quantum}%
  \BibitemOpen
  \bibfield  {author} {\bibinfo {author} {\bibfnamefont {D.~J.}\ \bibnamefont
  {Brod}}\ and\ \bibinfo {author} {\bibfnamefont {M.}~\bibnamefont
  {Oszmaniec}},\ }\href {\doibase 10.22331/q-2020-05-14-267} {\bibfield
  {journal} {\bibinfo  {journal} {Quantum}\ }\textbf {\bibinfo {volume} {4}},\
  \bibinfo {pages} {267} (\bibinfo {year} {2020})},\ \Eprint
  {http://arxiv.org/abs/1906.06696} {arXiv:1906.06696} \BibitemShut {NoStop}%
\bibitem [{\citenamefont {Renema}(2020)}]{renema_2020_Phys.Rev.A}%
  \BibitemOpen
  \bibfield  {author} {\bibinfo {author} {\bibfnamefont {J.~J.}\ \bibnamefont
  {Renema}},\ }\href {\doibase 10.1103/PhysRevA.101.063840} {\bibfield
  {journal} {\bibinfo  {journal} {Phys. Rev. A}\ }\textbf {\bibinfo {volume}
  {101}},\ \bibinfo {pages} {063840} (\bibinfo {year} {2020})}\BibitemShut
  {NoStop}%
\bibitem [{\citenamefont {Villalonga}\ \emph {et~al.}(2021)\citenamefont
  {Villalonga}, \citenamefont {Niu}, \citenamefont {Li}, \citenamefont {Neven},
  \citenamefont {Platt}, \citenamefont {Smelyanskiy},\ and\ \citenamefont
  {Boixo}}]{2109.11525}%
  \BibitemOpen
  \bibfield  {author} {\bibinfo {author} {\bibfnamefont {B.}~\bibnamefont
  {Villalonga}}, \bibinfo {author} {\bibfnamefont {M.~Y.}\ \bibnamefont {Niu}},
  \bibinfo {author} {\bibfnamefont {L.}~\bibnamefont {Li}}, \bibinfo {author}
  {\bibfnamefont {H.}~\bibnamefont {Neven}}, \bibinfo {author} {\bibfnamefont
  {J.~C.}\ \bibnamefont {Platt}}, \bibinfo {author} {\bibfnamefont {V.~N.}\
  \bibnamefont {Smelyanskiy}}, \ and\ \bibinfo {author} {\bibfnamefont
  {S.}~\bibnamefont {Boixo}},\ }\href@noop {} {\  (\bibinfo {year} {2021})},\
  \Eprint {http://arxiv.org/abs/arXiv:2109.11525} {arXiv:2109.11525}
  \BibitemShut {NoStop}%
\bibitem [{\citenamefont {Shi}\ and\ \citenamefont
  {Byrnes}(2021)}]{shi_2021_ArXiv}%
  \BibitemOpen
  \bibfield  {author} {\bibinfo {author} {\bibfnamefont {J.}~\bibnamefont
  {Shi}}\ and\ \bibinfo {author} {\bibfnamefont {T.}~\bibnamefont {Byrnes}},\
  }\href@noop {} {\bibfield  {journal} {\bibinfo  {journal} {arXiv:2105.09583
  [quant-ph]}\ } (\bibinfo {year} {2021})},\ \Eprint
  {http://arxiv.org/abs/2105.09583} {arXiv:2105.09583 [quant-ph]} \BibitemShut
  {NoStop}%
\bibitem [{\citenamefont {Pan}\ and\ \citenamefont
  {Zhang}(2021)}]{pan_2021_ArXiv}%
  \BibitemOpen
  \bibfield  {author} {\bibinfo {author} {\bibfnamefont {F.}~\bibnamefont
  {Pan}}\ and\ \bibinfo {author} {\bibfnamefont {P.}~\bibnamefont {Zhang}},\
  }\href@noop {} {\bibfield  {journal} {\bibinfo  {journal} {arXiv:2103.03074
  [physics, physics:quant-ph]}\ } (\bibinfo {year} {2021})},\ \Eprint
  {http://arxiv.org/abs/2103.03074} {arXiv:2103.03074 [physics,
  physics:quant-ph]} \BibitemShut {NoStop}%
\bibitem [{\citenamefont {Bulmer}\ \emph {et~al.}(2021)\citenamefont {Bulmer},
  \citenamefont {Bell}, \citenamefont {Chadwick}, \citenamefont {Jones},
  \citenamefont {Moise}, \citenamefont {Rigazzi}, \citenamefont {Thorbecke},
  \citenamefont {Haus}, \citenamefont {Van~Vaerenbergh}, \citenamefont {Patel},
  \citenamefont {Walmsley},\ and\ \citenamefont {Laing}}]{bulmer_2021_ArXiv}%
  \BibitemOpen
  \bibfield  {author} {\bibinfo {author} {\bibfnamefont {J.~F.~F.}\
  \bibnamefont {Bulmer}}, \bibinfo {author} {\bibfnamefont {B.~A.}\
  \bibnamefont {Bell}}, \bibinfo {author} {\bibfnamefont {R.~S.}\ \bibnamefont
  {Chadwick}}, \bibinfo {author} {\bibfnamefont {A.~E.}\ \bibnamefont {Jones}},
  \bibinfo {author} {\bibfnamefont {D.}~\bibnamefont {Moise}}, \bibinfo
  {author} {\bibfnamefont {A.}~\bibnamefont {Rigazzi}}, \bibinfo {author}
  {\bibfnamefont {J.}~\bibnamefont {Thorbecke}}, \bibinfo {author}
  {\bibfnamefont {U.-U.}\ \bibnamefont {Haus}}, \bibinfo {author}
  {\bibfnamefont {T.}~\bibnamefont {Van~Vaerenbergh}}, \bibinfo {author}
  {\bibfnamefont {R.~B.}\ \bibnamefont {Patel}}, \bibinfo {author}
  {\bibfnamefont {I.~A.}\ \bibnamefont {Walmsley}}, \ and\ \bibinfo {author}
  {\bibfnamefont {A.}~\bibnamefont {Laing}},\ }\href@noop {} {\bibfield
  {journal} {\bibinfo  {journal} {arXiv:2108.01622 [quant-ph]}\ } (\bibinfo
  {year} {2021})},\ \Eprint {http://arxiv.org/abs/2108.01622} {arXiv:2108.01622
  [quant-ph]} \BibitemShut {NoStop}%
\bibitem [{\citenamefont {Garc{\'{\i}}a-Patr{\'{o}}n}\ \emph
  {et~al.}(2019)\citenamefont {Garc{\'{\i}}a-Patr{\'{o}}n}, \citenamefont
  {Renema},\ and\ \citenamefont {Shchesnovich}}]{GarcaPatrn2019}%
  \BibitemOpen
  \bibfield  {author} {\bibinfo {author} {\bibfnamefont {R.}~\bibnamefont
  {Garc{\'{\i}}a-Patr{\'{o}}n}}, \bibinfo {author} {\bibfnamefont {J.~J.}\
  \bibnamefont {Renema}}, \ and\ \bibinfo {author} {\bibfnamefont
  {V.}~\bibnamefont {Shchesnovich}},\ }\href {\doibase
  10.22331/q-2019-08-05-169} {\bibfield  {journal} {\bibinfo  {journal}
  {Quantum}\ }\textbf {\bibinfo {volume} {3}},\ \bibinfo {pages} {169}
  (\bibinfo {year} {2019})}\BibitemShut {NoStop}%
\bibitem [{\citenamefont {Popova}\ and\ \citenamefont
  {Rubtsov}(2021)}]{popova_2021_ArXiv}%
  \BibitemOpen
  \bibfield  {author} {\bibinfo {author} {\bibfnamefont {A.~S.}\ \bibnamefont
  {Popova}}\ and\ \bibinfo {author} {\bibfnamefont {A.~N.}\ \bibnamefont
  {Rubtsov}},\ }\href@noop {} {\bibfield  {journal} {\bibinfo  {journal}
  {arXiv:2106.01445 [quant-ph]}\ } (\bibinfo {year} {2021})},\ \Eprint
  {http://arxiv.org/abs/2106.01445} {arXiv:2106.01445 [quant-ph]} \BibitemShut
  {NoStop}%
\bibitem [{\citenamefont
  {Shchesnovich}(2020)}]{shchesnovich_2020_Int.J.QuantumInform.}%
  \BibitemOpen
  \bibfield  {author} {\bibinfo {author} {\bibfnamefont {V.~S.}\ \bibnamefont
  {Shchesnovich}},\ }\href {\doibase 10.1142/S0219749920500446} {\bibfield
  {journal} {\bibinfo  {journal} {Int. J. Quantum Inform.}\ }\textbf {\bibinfo
  {volume} {18}},\ \bibinfo {pages} {2050044} (\bibinfo {year}
  {2020})}\BibitemShut {NoStop}%
\bibitem [{\citenamefont {Mosk}\ \emph {et~al.}(2012)\citenamefont {Mosk},
  \citenamefont {Lagendijk}, \citenamefont {Lerosey},\ and\ \citenamefont
  {Fink}}]{mosk_2012_NatPhoton}%
  \BibitemOpen
  \bibfield  {author} {\bibinfo {author} {\bibfnamefont {A.~P.}\ \bibnamefont
  {Mosk}}, \bibinfo {author} {\bibfnamefont {A.}~\bibnamefont {Lagendijk}},
  \bibinfo {author} {\bibfnamefont {G.}~\bibnamefont {Lerosey}}, \ and\
  \bibinfo {author} {\bibfnamefont {M.}~\bibnamefont {Fink}},\ }\href@noop {}
  {\bibfield  {journal} {\bibinfo  {journal} {Nat Photon}\ }\textbf {\bibinfo
  {volume} {6}},\ \bibinfo {pages} {283} (\bibinfo {year} {2012})}\BibitemShut
  {NoStop}%
\bibitem [{\citenamefont {Barak}\ and\ \citenamefont
  {{Ben-Aryeh}}(2007)}]{barak_2007_J.Opt.Soc.Am.B}%
  \BibitemOpen
  \bibfield  {author} {\bibinfo {author} {\bibfnamefont {R.}~\bibnamefont
  {Barak}}\ and\ \bibinfo {author} {\bibfnamefont {Y.}~\bibnamefont
  {{Ben-Aryeh}}},\ }\href {\doibase 10.1364/JOSAB.24.000231} {\bibfield
  {journal} {\bibinfo  {journal} {J. Opt. Soc. Am. B}\ }\textbf {\bibinfo
  {volume} {24}},\ \bibinfo {pages} {231} (\bibinfo {year} {2007})}\BibitemShut
  {NoStop}%
\bibitem [{\citenamefont {Crespi}\ \emph {et~al.}(2016)\citenamefont {Crespi},
  \citenamefont {Osellame}, \citenamefont {Ramponi}, \citenamefont
  {Bentivegna}, \citenamefont {Flamini}, \citenamefont {Spagnolo},
  \citenamefont {Viggianiello}, \citenamefont {Innocenti}, \citenamefont
  {Mataloni},\ and\ \citenamefont {Sciarrino}}]{crespi_2016_NatCommun}%
  \BibitemOpen
  \bibfield  {author} {\bibinfo {author} {\bibfnamefont {A.}~\bibnamefont
  {Crespi}}, \bibinfo {author} {\bibfnamefont {R.}~\bibnamefont {Osellame}},
  \bibinfo {author} {\bibfnamefont {R.}~\bibnamefont {Ramponi}}, \bibinfo
  {author} {\bibfnamefont {M.}~\bibnamefont {Bentivegna}}, \bibinfo {author}
  {\bibfnamefont {F.}~\bibnamefont {Flamini}}, \bibinfo {author} {\bibfnamefont
  {N.}~\bibnamefont {Spagnolo}}, \bibinfo {author} {\bibfnamefont
  {N.}~\bibnamefont {Viggianiello}}, \bibinfo {author} {\bibfnamefont
  {L.}~\bibnamefont {Innocenti}}, \bibinfo {author} {\bibfnamefont
  {P.}~\bibnamefont {Mataloni}}, \ and\ \bibinfo {author} {\bibfnamefont
  {F.}~\bibnamefont {Sciarrino}},\ }\href {\doibase 10.1038/ncomms10469}
  {\bibfield  {journal} {\bibinfo  {journal} {Nat Commun}\ }\textbf {\bibinfo
  {volume} {7}},\ \bibinfo {pages} {10469} (\bibinfo {year}
  {2016})}\BibitemShut {NoStop}%
\bibitem [{\citenamefont {Vellekoop}\ and\ \citenamefont
  {Mosk}(2007)}]{vellekoop_2007_OptLett}%
  \BibitemOpen
  \bibfield  {author} {\bibinfo {author} {\bibfnamefont {I.~M.}\ \bibnamefont
  {Vellekoop}}\ and\ \bibinfo {author} {\bibfnamefont {A.~P.}\ \bibnamefont
  {Mosk}},\ }\href {\doibase 10.1364/OL.32.002309} {\bibfield  {journal}
  {\bibinfo  {journal} {Opt. Lett.}\ }\textbf {\bibinfo {volume} {32}},\
  \bibinfo {pages} {2309} (\bibinfo {year} {2007})}\BibitemShut {NoStop}%
\bibitem [{\citenamefont {Rotter}\ and\ \citenamefont
  {Gigan}(2017)}]{rotter_2017_Rev.Mod.Phys.}%
  \BibitemOpen
  \bibfield  {author} {\bibinfo {author} {\bibfnamefont {S.}~\bibnamefont
  {Rotter}}\ and\ \bibinfo {author} {\bibfnamefont {S.}~\bibnamefont {Gigan}},\
  }\href {\doibase 10.1103/RevModPhys.89.015005} {\bibfield  {journal}
  {\bibinfo  {journal} {Rev. Mod. Phys.}\ }\textbf {\bibinfo {volume} {89}},\
  \bibinfo {pages} {015005} (\bibinfo {year} {2017})}\BibitemShut {NoStop}%
\bibitem [{\citenamefont {Beenakker}(2009)}]{beenakker_2009_RMT}%
  \BibitemOpen
  \bibfield  {author} {\bibinfo {author} {\bibfnamefont {C.~W.~J.}\
  \bibnamefont {Beenakker}},\ }\href@noop {} {\bibfield  {journal} {\bibinfo
  {journal} {arXiv:0904.1432 [cond-mat, physics:physics]}\ } (\bibinfo {year}
  {2009})},\ \Eprint {http://arxiv.org/abs/0904.1432} {arXiv:0904.1432
  [cond-mat, physics:physics]} \BibitemShut {NoStop}%
\bibitem [{\citenamefont {Goodman}(2007)}]{goodman2007speckle}%
  \BibitemOpen
  \bibfield  {author} {\bibinfo {author} {\bibfnamefont {J.}~\bibnamefont
  {Goodman}},\ }\href@noop {} {\emph {\bibinfo {title} {Speckle Phenomena in
  Optics: {{Theory}} and Applications}}}\ (\bibinfo  {publisher} {{Roberts \&
  Company}},\ \bibinfo {year} {2007})\BibitemShut {NoStop}%
\bibitem [{\citenamefont {Di~Lorenzo~Pires}\ \emph {et~al.}(2012)\citenamefont
  {Di~Lorenzo~Pires}, \citenamefont {Woudenberg},\ and\ \citenamefont {{van
  Exter}}}]{dilorenzopires_2012_Phys.Rev.A}%
  \BibitemOpen
  \bibfield  {author} {\bibinfo {author} {\bibfnamefont {H.}~\bibnamefont
  {Di~Lorenzo~Pires}}, \bibinfo {author} {\bibfnamefont {J.}~\bibnamefont
  {Woudenberg}}, \ and\ \bibinfo {author} {\bibfnamefont {M.~P.}\ \bibnamefont
  {{van Exter}}},\ }\href {\doibase 10.1103/PhysRevA.85.033807} {\bibfield
  {journal} {\bibinfo  {journal} {Phys. Rev. A}\ }\textbf {\bibinfo {volume}
  {85}},\ \bibinfo {pages} {033807} (\bibinfo {year} {2012})}\BibitemShut
  {NoStop}%
\bibitem [{\citenamefont {Defienne}\ \emph {et~al.}(2014)\citenamefont
  {Defienne}, \citenamefont {Barbieri}, \citenamefont {Chalopin}, \citenamefont
  {Chatel}, \citenamefont {Walmsley}, \citenamefont {Smith},\ and\
  \citenamefont {Gigan}}]{defienne_2014_Opt.Lett.a}%
  \BibitemOpen
  \bibfield  {author} {\bibinfo {author} {\bibfnamefont {H.}~\bibnamefont
  {Defienne}}, \bibinfo {author} {\bibfnamefont {M.}~\bibnamefont {Barbieri}},
  \bibinfo {author} {\bibfnamefont {B.}~\bibnamefont {Chalopin}}, \bibinfo
  {author} {\bibfnamefont {B.}~\bibnamefont {Chatel}}, \bibinfo {author}
  {\bibfnamefont {I.~A.}\ \bibnamefont {Walmsley}}, \bibinfo {author}
  {\bibfnamefont {B.~J.}\ \bibnamefont {Smith}}, \ and\ \bibinfo {author}
  {\bibfnamefont {S.}~\bibnamefont {Gigan}},\ }\href {\doibase
  10.1364/OL.39.006090} {\bibfield  {journal} {\bibinfo  {journal} {Optics
  Letters}\ }\textbf {\bibinfo {volume} {39}},\ \bibinfo {pages} {6090}
  (\bibinfo {year} {2014})}\BibitemShut {NoStop}%
\bibitem [{\citenamefont {Huisman}\ \emph {et~al.}(2014)\citenamefont
  {Huisman}, \citenamefont {Huisman}, \citenamefont {Mosk},\ and\ \citenamefont
  {Pinkse}}]{huisman_2014_ApplPhysB}%
  \BibitemOpen
  \bibfield  {author} {\bibinfo {author} {\bibfnamefont {T.~J.}\ \bibnamefont
  {Huisman}}, \bibinfo {author} {\bibfnamefont {S.~R.}\ \bibnamefont
  {Huisman}}, \bibinfo {author} {\bibfnamefont {A.~P.}\ \bibnamefont {Mosk}}, \
  and\ \bibinfo {author} {\bibfnamefont {P.~W.~H.}\ \bibnamefont {Pinkse}},\
  }\href {\doibase 10.1007/s00340-013-5742-5} {\bibfield  {journal} {\bibinfo
  {journal} {Appl. Phys. B}\ }\textbf {\bibinfo {volume} {116}},\ \bibinfo
  {pages} {603} (\bibinfo {year} {2014})}\BibitemShut {NoStop}%
\bibitem [{\citenamefont {Wolterink}\ \emph {et~al.}(2016)\citenamefont
  {Wolterink}, \citenamefont {Uppu}, \citenamefont {Ctistis}, \citenamefont
  {Vos}, \citenamefont {Boller},\ and\ \citenamefont
  {Pinkse}}]{wolterink_2016_PhysRevA}%
  \BibitemOpen
  \bibfield  {author} {\bibinfo {author} {\bibfnamefont {T.~A.~W.}\
  \bibnamefont {Wolterink}}, \bibinfo {author} {\bibfnamefont {R.}~\bibnamefont
  {Uppu}}, \bibinfo {author} {\bibfnamefont {G.}~\bibnamefont {Ctistis}},
  \bibinfo {author} {\bibfnamefont {W.~L.}\ \bibnamefont {Vos}}, \bibinfo
  {author} {\bibfnamefont {K.-J.}\ \bibnamefont {Boller}}, \ and\ \bibinfo
  {author} {\bibfnamefont {P.~W.~H.}\ \bibnamefont {Pinkse}},\ }\href {\doibase
  10.1103/PhysRevA.93.053817} {\bibfield  {journal} {\bibinfo  {journal} {Phys.
  Rev. A}\ }\textbf {\bibinfo {volume} {93}},\ \bibinfo {pages} {053817}
  (\bibinfo {year} {2016})}\BibitemShut {NoStop}%
\bibitem [{\citenamefont {Peeters}\ \emph {et~al.}(2010)\citenamefont
  {Peeters}, \citenamefont {Moerman},\ and\ \citenamefont {{van
  Exter}}}]{peeters_2010_Phys.Rev.Lett.}%
  \BibitemOpen
  \bibfield  {author} {\bibinfo {author} {\bibfnamefont {W.~H.}\ \bibnamefont
  {Peeters}}, \bibinfo {author} {\bibfnamefont {J.~J.~D.}\ \bibnamefont
  {Moerman}}, \ and\ \bibinfo {author} {\bibfnamefont {M.~P.}\ \bibnamefont
  {{van Exter}}},\ }\href {\doibase 10.1103/PhysRevLett.104.173601} {\bibfield
  {journal} {\bibinfo  {journal} {Phys. Rev. Lett.}\ }\textbf {\bibinfo
  {volume} {104}},\ \bibinfo {pages} {173601} (\bibinfo {year}
  {2010})}\BibitemShut {NoStop}%
\bibitem [{\citenamefont {Brod}\ \emph {et~al.}(2019)\citenamefont {Brod},
  \citenamefont {Galv{\~a}o}, \citenamefont {Crespi}, \citenamefont {Osellame},
  \citenamefont {Spagnolo},\ and\ \citenamefont {Sciarrino}}]{brod_2019_AP}%
  \BibitemOpen
  \bibfield  {author} {\bibinfo {author} {\bibfnamefont {D.~J.}\ \bibnamefont
  {Brod}}, \bibinfo {author} {\bibfnamefont {E.~F.}\ \bibnamefont
  {Galv{\~a}o}}, \bibinfo {author} {\bibfnamefont {A.}~\bibnamefont {Crespi}},
  \bibinfo {author} {\bibfnamefont {R.}~\bibnamefont {Osellame}}, \bibinfo
  {author} {\bibfnamefont {N.}~\bibnamefont {Spagnolo}}, \ and\ \bibinfo
  {author} {\bibfnamefont {F.}~\bibnamefont {Sciarrino}},\ }\href {\doibase
  10.1117/1.AP.1.3.034001} {\bibfield  {journal} {\bibinfo  {journal} {AP}\
  }\textbf {\bibinfo {volume} {1}},\ \bibinfo {pages} {034001} (\bibinfo {year}
  {2019})}\BibitemShut {NoStop}%
\bibitem [{\citenamefont {Scheel}(2004)}]{scheel_2004_ArXiv}%
  \BibitemOpen
  \bibfield  {author} {\bibinfo {author} {\bibfnamefont {S.}~\bibnamefont
  {Scheel}},\ }\href@noop {} {\bibfield  {journal} {\bibinfo  {journal}
  {arXiv:quant-ph/0406127}\ } (\bibinfo {year} {2004})},\ \Eprint
  {http://arxiv.org/abs/quant-ph/0406127} {arXiv:quant-ph/0406127} \BibitemShut
  {NoStop}%
\bibitem [{\citenamefont {Arkhipov}\ and\ \citenamefont
  {Kuperberg}(2012)}]{arkhipov_2012}%
  \BibitemOpen
  \bibfield  {author} {\bibinfo {author} {\bibfnamefont {A.}~\bibnamefont
  {Arkhipov}}\ and\ \bibinfo {author} {\bibfnamefont {G.}~\bibnamefont
  {Kuperberg}},\ }\href@noop {} {\bibfield  {journal} {\bibinfo  {journal}
  {Geom. Topol. Monogr.}\ }\textbf {\bibinfo {volume} {18}},\ \bibinfo {pages}
  {1} (\bibinfo {year} {2012})}\BibitemShut {NoStop}%
\bibitem [{\citenamefont {Bentkus}(2005)}]{bentkus_2005_TheoryProbab.Appl.}%
  \BibitemOpen
  \bibfield  {author} {\bibinfo {author} {\bibfnamefont {V.}~\bibnamefont
  {Bentkus}},\ }\href {\doibase 10.1137/S0040585X97981123} {\bibfield
  {journal} {\bibinfo  {journal} {Theory Probab. Appl.}\ }\textbf {\bibinfo
  {volume} {49}},\ \bibinfo {pages} {311} (\bibinfo {year} {2005})}\BibitemShut
  {NoStop}%
\bibitem [{\citenamefont {Meckes}(2009)}]{meckes_2009_}%
  \BibitemOpen
  \bibfield  {author} {\bibinfo {author} {\bibfnamefont {M.~W.}\ \bibnamefont
  {Meckes}},\ }\href {\doibase 10.1214/09-IMSCOLL514} {\emph {\bibinfo {title}
  {Some Results on Random Circulant Matrices}}}\ (\bibinfo  {publisher}
  {{Institute of Mathematical Statistics}},\ \bibinfo {year} {2009})\ pp.\
  \bibinfo {pages} {213--223}\BibitemShut {NoStop}%
\bibitem [{\citenamefont {Mastrodonato}\ and\ \citenamefont
  {Tumulka}(2007)}]{mastrodonato_2007_LettMathPhys}%
  \BibitemOpen
  \bibfield  {author} {\bibinfo {author} {\bibfnamefont {C.}~\bibnamefont
  {Mastrodonato}}\ and\ \bibinfo {author} {\bibfnamefont {R.}~\bibnamefont
  {Tumulka}},\ }\href {\doibase 10.1007/s11005-007-0194-7} {\bibfield
  {journal} {\bibinfo  {journal} {Lett Math Phys}\ }\textbf {\bibinfo {volume}
  {82}},\ \bibinfo {pages} {51} (\bibinfo {year} {2007})}\BibitemShut {NoStop}%
\bibitem [{\citenamefont {Petz}\ and\ \citenamefont
  {R{\'e}ffy}(2005{\natexlab{a}})}]{petz_2005_Period.Math.Hung.}%
  \BibitemOpen
  \bibfield  {author} {\bibinfo {author} {\bibfnamefont {D.}~\bibnamefont
  {Petz}}\ and\ \bibinfo {author} {\bibfnamefont {J.}~\bibnamefont
  {R{\'e}ffy}},\ }\href {\doibase 10.1023/b:mahu.0000040542.56072.ab}
  {\bibfield  {journal} {\bibinfo  {journal} {Periodica Mathematica Hungarica}\
  }\textbf {\bibinfo {volume} {49}},\ \bibinfo {pages} {103} (\bibinfo {year}
  {2005}{\natexlab{a}})}\BibitemShut {NoStop}%
\bibitem [{\citenamefont {Petz}\ and\ \citenamefont
  {R{\'e}ffy}(2005{\natexlab{b}})}]{petz_2005_Probab.TheoryRelat.Fields}%
  \BibitemOpen
  \bibfield  {author} {\bibinfo {author} {\bibfnamefont {D.}~\bibnamefont
  {Petz}}\ and\ \bibinfo {author} {\bibfnamefont {J.}~\bibnamefont
  {R{\'e}ffy}},\ }\href {\doibase 10.1007/s00440-004-0420-5} {\bibfield
  {journal} {\bibinfo  {journal} {Probab. Theory Relat. Fields}\ }\textbf
  {\bibinfo {volume} {133}},\ \bibinfo {pages} {175} (\bibinfo {year}
  {2005}{\natexlab{b}})}\BibitemShut {NoStop}%
\bibitem [{\citenamefont {R{\'e}ffy}(2005)}]{reffy_2005_undefined}%
  \BibitemOpen
  \bibfield  {author} {\bibinfo {author} {\bibfnamefont {J.}~\bibnamefont
  {R{\'e}ffy}},\ }\emph {\bibinfo {title} {Asymptotics of Random Unitaries}},\
  \href@noop {} {Ph.D. thesis},\ \bibinfo  {school} {Budapest University of
  Technology and Economics}, \bibinfo {address} {{Budapest}} (\bibinfo {year}
  {2005})\BibitemShut {NoStop}%
\bibitem [{\citenamefont {Nielsen}\ and\ \citenamefont
  {Chuang}(2010)}]{nielsen2010quantum}%
  \BibitemOpen
  \bibfield  {author} {\bibinfo {author} {\bibfnamefont {M.}~\bibnamefont
  {Nielsen}}\ and\ \bibinfo {author} {\bibfnamefont {I.}~\bibnamefont
  {Chuang}},\ }\href@noop {} {\emph {\bibinfo {title} {Quantum Computation and
  Quantum Information: 10th Anniversary Edition}}}\ (\bibinfo  {publisher}
  {{Cambridge University Press}},\ \bibinfo {year} {2010})\BibitemShut
  {NoStop}%
\bibitem [{\citenamefont {Lita}\ \emph {et~al.}(2008)\citenamefont {Lita},
  \citenamefont {Miller},\ and\ \citenamefont {Nam}}]{lita_2008_OptExpress}%
  \BibitemOpen
  \bibfield  {author} {\bibinfo {author} {\bibfnamefont {A.~E.}\ \bibnamefont
  {Lita}}, \bibinfo {author} {\bibfnamefont {A.~J.}\ \bibnamefont {Miller}}, \
  and\ \bibinfo {author} {\bibfnamefont {S.~W.}\ \bibnamefont {Nam}},\ }\href
  {\doibase 10.1364/OE.16.003032} {\bibfield  {journal} {\bibinfo  {journal}
  {Opt. Express}\ }\textbf {\bibinfo {volume} {16}},\ \bibinfo {pages} {3032}
  (\bibinfo {year} {2008})}\BibitemShut {NoStop}%
\bibitem [{\citenamefont {Reddy}\ \emph {et~al.}(2019)\citenamefont {Reddy},
  \citenamefont {Reddy}, \citenamefont {Lita}, \citenamefont {Nam},
  \citenamefont {Mirin},\ and\ \citenamefont {Verma}}]{reddy_2019}%
  \BibitemOpen
  \bibfield  {author} {\bibinfo {author} {\bibfnamefont {D.~V.}\ \bibnamefont
  {Reddy}}, \bibinfo {author} {\bibfnamefont {D.~V.}\ \bibnamefont {Reddy}},
  \bibinfo {author} {\bibfnamefont {A.~E.}\ \bibnamefont {Lita}}, \bibinfo
  {author} {\bibfnamefont {S.~W.}\ \bibnamefont {Nam}}, \bibinfo {author}
  {\bibfnamefont {R.~P.}\ \bibnamefont {Mirin}}, \ and\ \bibinfo {author}
  {\bibfnamefont {V.~B.}\ \bibnamefont {Verma}},\ }in\ \href {\doibase
  10.1364/CQO.2019.W2B.2} {\emph {\bibinfo {booktitle} {Rochester
  {{Conference}} on {{Coherence}} and {{Quantum Optics}} ({{CQO}}-11) (2019),
  Paper {{W2B}}.2}}}\ (\bibinfo  {publisher} {{Optical Society of America}},\
  \bibinfo {year} {2019})\ p.\ \bibinfo {pages} {W2B.2}\BibitemShut {NoStop}%
\bibitem [{\citenamefont {Doerner}\ \emph {et~al.}(2017)\citenamefont
  {Doerner}, \citenamefont {Kuzmin}, \citenamefont {Wuensch}, \citenamefont
  {Charaev}, \citenamefont {Boes}, \citenamefont {Zwick},\ and\ \citenamefont
  {Siegel}}]{doerner_2017_Appl.Phys.Lett.}%
  \BibitemOpen
  \bibfield  {author} {\bibinfo {author} {\bibfnamefont {S.}~\bibnamefont
  {Doerner}}, \bibinfo {author} {\bibfnamefont {A.}~\bibnamefont {Kuzmin}},
  \bibinfo {author} {\bibfnamefont {S.}~\bibnamefont {Wuensch}}, \bibinfo
  {author} {\bibfnamefont {I.}~\bibnamefont {Charaev}}, \bibinfo {author}
  {\bibfnamefont {F.}~\bibnamefont {Boes}}, \bibinfo {author} {\bibfnamefont
  {T.}~\bibnamefont {Zwick}}, \ and\ \bibinfo {author} {\bibfnamefont
  {M.}~\bibnamefont {Siegel}},\ }\href {\doibase 10.1063/1.4993779} {\bibfield
  {journal} {\bibinfo  {journal} {Appl. Phys. Lett.}\ }\textbf {\bibinfo
  {volume} {111}},\ \bibinfo {pages} {032603} (\bibinfo {year}
  {2017})}\BibitemShut {NoStop}%
\bibitem [{\citenamefont {Gaggero}\ \emph {et~al.}(2019)\citenamefont
  {Gaggero}, \citenamefont {Martini}, \citenamefont {Mattioli}, \citenamefont
  {Chiarello}, \citenamefont {Cernansky}, \citenamefont {Politi},\ and\
  \citenamefont {Leoni}}]{gaggero_2019_Optica}%
  \BibitemOpen
  \bibfield  {author} {\bibinfo {author} {\bibfnamefont {A.}~\bibnamefont
  {Gaggero}}, \bibinfo {author} {\bibfnamefont {F.}~\bibnamefont {Martini}},
  \bibinfo {author} {\bibfnamefont {F.}~\bibnamefont {Mattioli}}, \bibinfo
  {author} {\bibfnamefont {F.}~\bibnamefont {Chiarello}}, \bibinfo {author}
  {\bibfnamefont {R.}~\bibnamefont {Cernansky}}, \bibinfo {author}
  {\bibfnamefont {A.}~\bibnamefont {Politi}}, \ and\ \bibinfo {author}
  {\bibfnamefont {R.}~\bibnamefont {Leoni}},\ }\href {\doibase
  10.1364/OPTICA.6.000823} {\bibfield  {journal} {\bibinfo  {journal} {Optica,
  OPTICA}\ }\textbf {\bibinfo {volume} {6}},\ \bibinfo {pages} {823} (\bibinfo
  {year} {2019})}\BibitemShut {NoStop}%
\bibitem [{\citenamefont {Miki}\ \emph {et~al.}(2014)\citenamefont {Miki},
  \citenamefont {Yamashita}, \citenamefont {Wang},\ and\ \citenamefont
  {Terai}}]{miki_2014_Opt.ExpressOE}%
  \BibitemOpen
  \bibfield  {author} {\bibinfo {author} {\bibfnamefont {S.}~\bibnamefont
  {Miki}}, \bibinfo {author} {\bibfnamefont {T.}~\bibnamefont {Yamashita}},
  \bibinfo {author} {\bibfnamefont {Z.}~\bibnamefont {Wang}}, \ and\ \bibinfo
  {author} {\bibfnamefont {H.}~\bibnamefont {Terai}},\ }\href {\doibase
  10.1364/OE.22.007811} {\bibfield  {journal} {\bibinfo  {journal} {Opt.
  Express, OE}\ }\textbf {\bibinfo {volume} {22}},\ \bibinfo {pages} {7811}
  (\bibinfo {year} {2014})}\BibitemShut {NoStop}%
\bibitem [{\citenamefont {Miyajima}\ \emph {et~al.}(2018)\citenamefont
  {Miyajima}, \citenamefont {Yabuno}, \citenamefont {Miki}, \citenamefont
  {Yamashita},\ and\ \citenamefont {Terai}}]{miyajima_2018_Opt.ExpressOE}%
  \BibitemOpen
  \bibfield  {author} {\bibinfo {author} {\bibfnamefont {S.}~\bibnamefont
  {Miyajima}}, \bibinfo {author} {\bibfnamefont {M.}~\bibnamefont {Yabuno}},
  \bibinfo {author} {\bibfnamefont {S.}~\bibnamefont {Miki}}, \bibinfo {author}
  {\bibfnamefont {T.}~\bibnamefont {Yamashita}}, \ and\ \bibinfo {author}
  {\bibfnamefont {H.}~\bibnamefont {Terai}},\ }\href {\doibase
  10.1364/OE.26.029045} {\bibfield  {journal} {\bibinfo  {journal} {Opt.
  Express, OE}\ }\textbf {\bibinfo {volume} {26}},\ \bibinfo {pages} {29045}
  (\bibinfo {year} {2018})}\BibitemShut {NoStop}%
\bibitem [{\citenamefont {Sinclair}\ \emph {et~al.}(2019)\citenamefont
  {Sinclair}, \citenamefont {Schroeder}, \citenamefont {Zhu}, \citenamefont
  {Colangelo}, \citenamefont {Glasby}, \citenamefont {Mauskopf}, \citenamefont
  {Mani},\ and\ \citenamefont
  {Berggren}}]{sinclair_2019_IEEETrans.Appl.Supercond.}%
  \BibitemOpen
  \bibfield  {author} {\bibinfo {author} {\bibfnamefont {A.~K.}\ \bibnamefont
  {Sinclair}}, \bibinfo {author} {\bibfnamefont {E.}~\bibnamefont {Schroeder}},
  \bibinfo {author} {\bibfnamefont {D.}~\bibnamefont {Zhu}}, \bibinfo {author}
  {\bibfnamefont {M.}~\bibnamefont {Colangelo}}, \bibinfo {author}
  {\bibfnamefont {J.}~\bibnamefont {Glasby}}, \bibinfo {author} {\bibfnamefont
  {P.~D.}\ \bibnamefont {Mauskopf}}, \bibinfo {author} {\bibfnamefont
  {H.}~\bibnamefont {Mani}}, \ and\ \bibinfo {author} {\bibfnamefont {K.~K.}\
  \bibnamefont {Berggren}},\ }\href {\doibase 10.1109/TASC.2019.2899329}
  {\bibfield  {journal} {\bibinfo  {journal} {IEEE Transactions on Applied
  Superconductivity}\ }\textbf {\bibinfo {volume} {29}},\ \bibinfo {pages} {1}
  (\bibinfo {year} {2019})}\BibitemShut {NoStop}%
\bibitem [{\citenamefont {Zhao}\ \emph {et~al.}(2013)\citenamefont {Zhao},
  \citenamefont {McCaughan}, \citenamefont {Bellei}, \citenamefont {Najafi},
  \citenamefont {De~Fazio}, \citenamefont {Dane}, \citenamefont {Ivry},\ and\
  \citenamefont {Berggren}}]{zhao_2013_Appl.Phys.Lett.}%
  \BibitemOpen
  \bibfield  {author} {\bibinfo {author} {\bibfnamefont {Q.}~\bibnamefont
  {Zhao}}, \bibinfo {author} {\bibfnamefont {A.}~\bibnamefont {McCaughan}},
  \bibinfo {author} {\bibfnamefont {F.}~\bibnamefont {Bellei}}, \bibinfo
  {author} {\bibfnamefont {F.}~\bibnamefont {Najafi}}, \bibinfo {author}
  {\bibfnamefont {D.}~\bibnamefont {De~Fazio}}, \bibinfo {author}
  {\bibfnamefont {A.}~\bibnamefont {Dane}}, \bibinfo {author} {\bibfnamefont
  {Y.}~\bibnamefont {Ivry}}, \ and\ \bibinfo {author} {\bibfnamefont {K.~K.}\
  \bibnamefont {Berggren}},\ }\href {\doibase 10.1063/1.4823542} {\bibfield
  {journal} {\bibinfo  {journal} {Appl. Phys. Lett.}\ }\textbf {\bibinfo
  {volume} {103}},\ \bibinfo {pages} {142602} (\bibinfo {year}
  {2013})}\BibitemShut {NoStop}%
\bibitem [{\citenamefont {Zhao}\ \emph {et~al.}(2017)\citenamefont {Zhao},
  \citenamefont {Zhu}, \citenamefont {Calandri}, \citenamefont {Dane},
  \citenamefont {McCaughan}, \citenamefont {Bellei}, \citenamefont {Wang},
  \citenamefont {Santavicca},\ and\ \citenamefont
  {Berggren}}]{zhao_2017_Nat.Photonics}%
  \BibitemOpen
  \bibfield  {author} {\bibinfo {author} {\bibfnamefont {Q.-Y.}\ \bibnamefont
  {Zhao}}, \bibinfo {author} {\bibfnamefont {D.}~\bibnamefont {Zhu}}, \bibinfo
  {author} {\bibfnamefont {N.}~\bibnamefont {Calandri}}, \bibinfo {author}
  {\bibfnamefont {A.~E.}\ \bibnamefont {Dane}}, \bibinfo {author}
  {\bibfnamefont {A.~N.}\ \bibnamefont {McCaughan}}, \bibinfo {author}
  {\bibfnamefont {F.}~\bibnamefont {Bellei}}, \bibinfo {author} {\bibfnamefont
  {H.-Z.}\ \bibnamefont {Wang}}, \bibinfo {author} {\bibfnamefont {D.~F.}\
  \bibnamefont {Santavicca}}, \ and\ \bibinfo {author} {\bibfnamefont {K.~K.}\
  \bibnamefont {Berggren}},\ }\href {\doibase 10.1038/nphoton.2017.35}
  {\bibfield  {journal} {\bibinfo  {journal} {Nature Photonics}\ }\textbf
  {\bibinfo {volume} {11}},\ \bibinfo {pages} {247} (\bibinfo {year}
  {2017})}\BibitemShut {NoStop}%
\bibitem [{\citenamefont {Zhu}\ \emph {et~al.}(2018)\citenamefont {Zhu},
  \citenamefont {Zhao}, \citenamefont {Choi}, \citenamefont {Lu}, \citenamefont
  {Dane}, \citenamefont {Englund},\ and\ \citenamefont
  {Berggren}}]{zhu_2018_Nat.Nanotechnol.}%
  \BibitemOpen
  \bibfield  {author} {\bibinfo {author} {\bibfnamefont {D.}~\bibnamefont
  {Zhu}}, \bibinfo {author} {\bibfnamefont {Q.-Y.}\ \bibnamefont {Zhao}},
  \bibinfo {author} {\bibfnamefont {H.}~\bibnamefont {Choi}}, \bibinfo {author}
  {\bibfnamefont {T.-J.}\ \bibnamefont {Lu}}, \bibinfo {author} {\bibfnamefont
  {A.~E.}\ \bibnamefont {Dane}}, \bibinfo {author} {\bibfnamefont
  {D.}~\bibnamefont {Englund}}, \ and\ \bibinfo {author} {\bibfnamefont
  {K.~K.}\ \bibnamefont {Berggren}},\ }\href {\doibase
  10.1038/s41565-018-0160-9} {\bibfield  {journal} {\bibinfo  {journal} {Nature
  Nanotechnology}\ }\textbf {\bibinfo {volume} {13}},\ \bibinfo {pages} {596}
  (\bibinfo {year} {2018})}\BibitemShut {NoStop}%
\bibitem [{\citenamefont {Allmaras}\ \emph {et~al.}(2020)\citenamefont
  {Allmaras}, \citenamefont {Wollman}, \citenamefont {Beyer}, \citenamefont
  {Briggs}, \citenamefont {Korzh}, \citenamefont {Bumble},\ and\ \citenamefont
  {Shaw}}]{allmaras_2020_NanoLett.}%
  \BibitemOpen
  \bibfield  {author} {\bibinfo {author} {\bibfnamefont {J.~P.}\ \bibnamefont
  {Allmaras}}, \bibinfo {author} {\bibfnamefont {E.~E.}\ \bibnamefont
  {Wollman}}, \bibinfo {author} {\bibfnamefont {A.~D.}\ \bibnamefont {Beyer}},
  \bibinfo {author} {\bibfnamefont {R.~M.}\ \bibnamefont {Briggs}}, \bibinfo
  {author} {\bibfnamefont {B.~A.}\ \bibnamefont {Korzh}}, \bibinfo {author}
  {\bibfnamefont {B.}~\bibnamefont {Bumble}}, \ and\ \bibinfo {author}
  {\bibfnamefont {M.~D.}\ \bibnamefont {Shaw}},\ }\href {\doibase
  10.1021/acs.nanolett.0c00246} {\bibfield  {journal} {\bibinfo  {journal}
  {Nano Letters}\ }\textbf {\bibinfo {volume} {20}},\ \bibinfo {pages} {2163}
  (\bibinfo {year} {2020})}\BibitemShut {NoStop}%
\bibitem [{\citenamefont {Wollman}\ \emph {et~al.}(2019)\citenamefont
  {Wollman}, \citenamefont {Wollman}, \citenamefont {Verma}, \citenamefont
  {Verma}, \citenamefont {Lita}, \citenamefont {Farr}, \citenamefont {Shaw},
  \citenamefont {Mirin},\ and\ \citenamefont
  {Nam}}]{wollman_2019_Opt.ExpressOE}%
  \BibitemOpen
  \bibfield  {author} {\bibinfo {author} {\bibfnamefont {E.~E.}\ \bibnamefont
  {Wollman}}, \bibinfo {author} {\bibfnamefont {E.~E.}\ \bibnamefont
  {Wollman}}, \bibinfo {author} {\bibfnamefont {V.~B.}\ \bibnamefont {Verma}},
  \bibinfo {author} {\bibfnamefont {V.~B.}\ \bibnamefont {Verma}}, \bibinfo
  {author} {\bibfnamefont {A.~E.}\ \bibnamefont {Lita}}, \bibinfo {author}
  {\bibfnamefont {W.~H.}\ \bibnamefont {Farr}}, \bibinfo {author}
  {\bibfnamefont {M.~D.}\ \bibnamefont {Shaw}}, \bibinfo {author}
  {\bibfnamefont {R.~P.}\ \bibnamefont {Mirin}}, \ and\ \bibinfo {author}
  {\bibfnamefont {S.~W.}\ \bibnamefont {Nam}},\ }\href {\doibase
  10.1364/OE.27.035279} {\bibfield  {journal} {\bibinfo  {journal} {Opt.
  Express, OE}\ }\textbf {\bibinfo {volume} {27}},\ \bibinfo {pages} {35279}
  (\bibinfo {year} {2019})}\BibitemShut {NoStop}%
\bibitem [{\citenamefont {Allman}\ \emph {et~al.}(2015)\citenamefont {Allman},
  \citenamefont {Verma}, \citenamefont {Stevens}, \citenamefont {Gerrits},
  \citenamefont {Horansky}, \citenamefont {Lita}, \citenamefont {Marsili},
  \citenamefont {Beyer}, \citenamefont {Shaw}, \citenamefont {Kumor},
  \citenamefont {Mirin},\ and\ \citenamefont
  {Nam}}]{allman_2015_Appl.Phys.Lett.}%
  \BibitemOpen
  \bibfield  {author} {\bibinfo {author} {\bibfnamefont {M.~S.}\ \bibnamefont
  {Allman}}, \bibinfo {author} {\bibfnamefont {V.~B.}\ \bibnamefont {Verma}},
  \bibinfo {author} {\bibfnamefont {M.}~\bibnamefont {Stevens}}, \bibinfo
  {author} {\bibfnamefont {T.}~\bibnamefont {Gerrits}}, \bibinfo {author}
  {\bibfnamefont {R.~D.}\ \bibnamefont {Horansky}}, \bibinfo {author}
  {\bibfnamefont {A.~E.}\ \bibnamefont {Lita}}, \bibinfo {author}
  {\bibfnamefont {F.}~\bibnamefont {Marsili}}, \bibinfo {author} {\bibfnamefont
  {A.}~\bibnamefont {Beyer}}, \bibinfo {author} {\bibfnamefont {M.~D.}\
  \bibnamefont {Shaw}}, \bibinfo {author} {\bibfnamefont {D.}~\bibnamefont
  {Kumor}}, \bibinfo {author} {\bibfnamefont {R.}~\bibnamefont {Mirin}}, \ and\
  \bibinfo {author} {\bibfnamefont {S.~W.}\ \bibnamefont {Nam}},\ }\href
  {\doibase 10.1063/1.4921318} {\bibfield  {journal} {\bibinfo  {journal}
  {Appl. Phys. Lett.}\ }\textbf {\bibinfo {volume} {106}},\ \bibinfo {pages}
  {192601} (\bibinfo {year} {2015})}\BibitemShut {NoStop}%
\end{thebibliography}%

\end{document}